\documentclass[useAMS,usenatbib,usegraphicx]{mn2e}
\usepackage{amssymb,amsmath,amsfonts,graphicx}
\usepackage{epsfig,color}

\usepackage{pgf}
\usepackage{bm}

\def\gsim { \lower .75ex \hbox{$\sim$} \llap{\raise .27ex \hbox{$>$}} }
\def\lsim { \lower .75ex \hbox{$\sim$} \llap{\raise .27ex \hbox{$<$}} }

\usepackage{subcaption}

\usepackage{times}
\usepackage[bookmarks,bookmarksnumbered,colorlinks=true, citecolor=blue, linkcolor=black]{hyperref}
\usepackage{color}

\begin{document}
\title[Stellar halos in Illustris]{Stellar Halos in Illustris: Probing the Histories of Milky Way-Mass Galaxies}

\author[Elias et al.]{
\parbox[t]{\textwidth}{
Lydia M. Elias$^{1}$$\thanks{E-mail: lydia.elias@email.ucr.edu}$,     
Laura V. Sales$^{1}$$\thanks{Hellman Fellow}$,
Peter Creasey$^{1}$,
Michael C. Cooper$^{2}$,
James S. Bullock$^{2}$,
Michael R. Rich$^{3}$ and
Lars Hernquist$^{4}$}
\\
\\
$^{1}$ Department of Physics and Astronomy, University of California Riverside, 900 University Ave., CA92507, US\\
$^{2}$Center for Cosmology, Department of Physics \& Astronomy, University of California, Irvine, 4129 Reines Hall, Irvine, CA 92697, USA \\
$^{3}$Dept. of Physics and Astronomy, UCLA, Los Angeles, CA 90095-1547\\
$^{4}$Harvard-Smithsonian Center for Astrophysics, 60 Garden Street, Cambridge, MA, 02138, USA\\}

\maketitle

\begin{abstract}

  The existence of stellar halos around galaxies is a natural
  prediction of the hierarchical nature of the $\Lambda$CDM model.
  Recent observations of Milky Way-like galaxies have revealed a wide
  range in stellar halo mass, including cases with no statistically
  significant detection of a stellar halo, as in the case of M101,
  NGC3351 and NGC1042. We use the Illustris simulation to investigate
  the scatter in stellar halo content and, in particular, to study the
  formation of galaxies with the smallest fraction of this diffuse
  component. Stellar halos are far from spherical, which diminishes
  the surface brightness of the stellar halo for face-on disks. Once accounting
  for projection effects, we find that the stellar halo fraction
  $f_{SH}$ correlates strongly with galaxy morphology and star
  formation rate, but not with environment, in agreement with
  observations. Galaxies with the lowest stellar halo fractions are
  disk-dominated, star-forming and assemble their dark matter halos on
  average earlier than galaxies with similar stellar masses. Accreted
  satellites are also lower in stellar mass and have earlier infall times 
  than centrals with high $f_{\rm SH}$. In situ rather than
  accreted stars dominate the stellar halos of galaxies with the
  lowest stellar halo fractions, with a transition radius from in situ
  to accretion-dominated $r \sim 45$ kpc.  Our results extrapolated to
  real galaxies such as M101 may indicate that these galaxies inhabit
  old halos which endured mergers only at higher redshifts and evolved
  relatively unperturbed in the last $\sim 10$ Gyrs.
\end{abstract}


\section{Introduction}
\label{sec:intro}

Stars can be found out to very large distances away from the center of
galaxies and certainly well beyond their optical radii. This extended
and diffuse stellar component --the stellar halo-- is typically
composed of metal poor stars that are often distributed
inhomogeneously and present substructures in both configuration as
well as in velocity space. This phase-space coherence was predicted
early by cosmological models of galaxy assembly
\citep{Johnston1996,Johnston1998,Helmi1999,Bullock2001} and also found
in observations of stellar moving groups in the Milky Way
\citep{Ibata2014,Ibata1994,Shapley1938,Wilman2005} as well as in
several tidal streams
\citep{Lynden-Bell1995,Belokurov2006,Grillmair2009}. In fact, within
the leading cosmological framework Lambda Cold Dark Matter
($\Lambda$CDM), stellar halos are the natural outcome of the galaxy
formation process where galaxies form hierarchically by aggregating
smaller ones \citep{WhiteRees1978,White1996}. The disruption of these
satellite galaxies deposits stellar debris in the outskirts of the
galactic domain of the host galaxy, building up and shaping the
stellar halo component.  Stellar halos carry information on the
properties of satellite galaxies that no longer exist and on the
assembly history of the host galaxy. Because they are extended (they
can reach $100$ kpc and beyond for $L_*$ galaxies), dynamical times
are long, and tidal features such as streams and shells --records of
past disruptions-- can survive for very long timescales
\citep{Helmi1999,Bullock2005}, making stellar halos invaluable proof
of the cosmological model.

The stellar halos of our own Milky Way and closest neighboring galaxy
M31 have been studied most extensively. In particular, our own Galaxy
offers an advantageous view from where to select halo stars with
precise photometry and kinematics either within the solar
neighbourhood \citep[by specific cuts in metallicity and or
kinematics, e.g. ][]{Eggen1962,Chiba2000,Helmi2017} or by looking to
stars that are beyond the confines of the disk
\citep{Morrison2000,Fernandez-Trincado2015}. Photometric campaigns
such as that of SDSS/SEGUE have played a fundamental role in building
our current understanding of the Galactic stellar halo by revealing a
rich level of substructure where streams and overdensities abound
\citep{Belokurov2006,Bell2008,Juric2008,Odenkirchen2001,Grillmair2006}. Deep
observations of our closest neighbour M31 by the PAndAS survey
\citep{McConnachie2009} also reveal a complex stellar halo structure
with a massive stream dominating most of the light, but with several other
substructures easily identifiable
\citep{Ibata2014,Ferguson2002,Tanaka2010}. An unexpected conclusion of
these studies is that the stellar halo of M31 is significantly more
massive than that of our own Galaxy: $4 \%$ compared to $\sim 1\%$,
with possible interpretations pointing to a more active merging
history for M31 than for the Milky Way. Stellar halos may therefore
also provide an avenue to reconstruct the past merger history of
galaxies once the link between the galaxy build-up and its stellar
halo structure, mass and shape is fully understood.

The large number of substructures seen in the Milky Way and Andromeda
 serves as confirmation that accretion events are responsible to
a large extent for the build up of our stellar halo. However,
kinematics and metallicity analysis of stars suggest a more
complex structure for the stellar halo where more than one component
is needed \citep{Carollo2007,Deason2013,Carollo2010,Tissera2014}, including
some hints that the inner stellar halos may have a significant
contribution from in situ stars \citep{Gilbert2014,Reitzel2002,Bonaca2017}, 
as opposed to the common view of an accreted origin for
stellar halos. The mechanisms able to propel in situ born stars onto
more external orbits are not yet well understood, nor has a convincing
way been demonstrated to observationally distinguish between these two possible origins
for halo stars in external galaxies.

Clearly, a complete understanding of stellar halos will come from
surveying more galaxies beyond the Local Group. Such observations are
challenging due to the associated low surface brightness in these
halos. For instance, if one were to place a Milky Way-like stellar
halo uniformly in a sphere of radius $100$ kpc it would have a stellar
surface density of $1.6 \times 10^4$ $M_\odot/\rm kpc^2$ in
projection, which means that we would need to reach
levels of $\sim$30.9 mag/arcsec$^2$ to detect it. Although this
calculation is only approximate (in the sense that stars will
distribute in some power law and not homogeneously and also in that it
ignores substructures), it provides a good intuition about the high
sensitivity needed in observations of extragalactic stellar halos.

A clever way to circumvent this limitation is by means of stacking
images of thousands galaxies and their halos \citep[e.g.,
][]{Zibetti2004,Tal2011,DSouza2014}.  Another possibility is detailed
observations of a handful of galaxies with HST, such as the approach
taken by the GHOSTS survey \citep{Radburn-Smith2011,Monachesi2016},
for which random fields across the halo are selected providing
information on localized regions with resolved stellar
populations. More recently, the availability of telescopes
with full time dedication have allowed the mapping of several stellar
halos by very long exposures. Such is the case of the Dragonfly telescope
\citep{Abraham2014,vanDokkum2014,Merritt2016} or the HERON survey
\citep{Rich2017} reaching an unprecedented depth up to $29$-$32$
mag/arcsec$^2$, in an extension of a technique used previously to
uncover several extragalactic streams
\citep{Martinez-Delgado2010,Tal2009}.

The emergent picture from all these recent studies, in agreement with the
already detected differences between the Milky Way and M31 halos, is a
wide diversity in the amount, shape and gradients of stellar halos
among different galaxies \citep{Monachesi2016,Harmsen2017,Merritt2016}. Even within the limited sample of deep observations
that is currently available, some intriguing cases have already
arisen. In particular, the apparent non-detection of M101's stellar
halo by \citet{vanDokkum2014} is at odds with the predictions from
$\Lambda$CDM. {\it How can a galaxy avoid merging and disrupting satellites
  throughout its entire history?}. Furthermore, \citet{Merritt2016}
present two additional cases, NGC1042 and NGC3351 together with
M101, among their eight galaxies with comparable stellar mass to the
Milky Way which are also consistent with no stellar
halo component down to $32$ mag/arcsec$^2$. This apparent lack of
stellar halo, if proven
not to be an observational artifact, seems not uncommon in the galaxy
population and is therefore an interesting puzzle to be explored within
the current models of galaxy formation.


\begin{center}
\begin{figure*}
\includegraphics[width=0.49 \linewidth,clip]{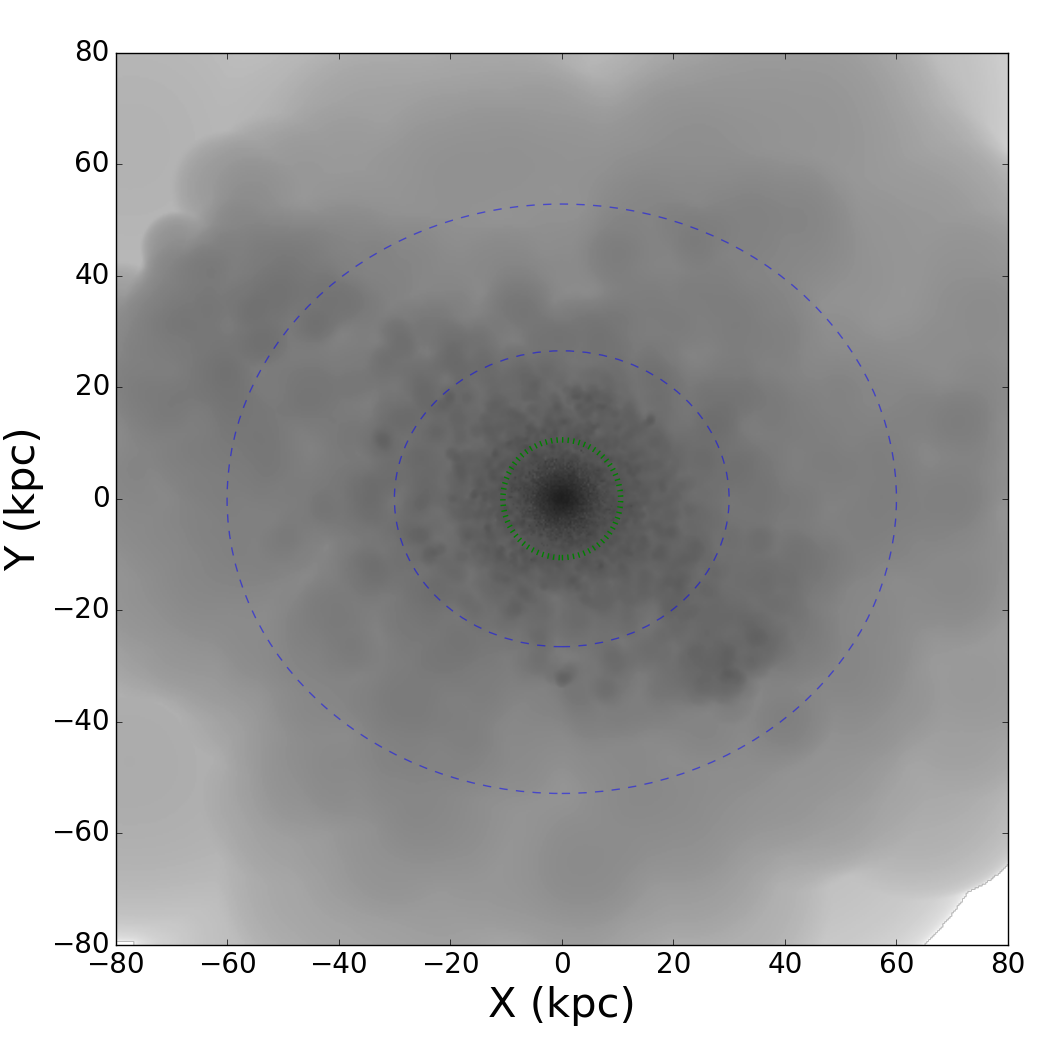}
\includegraphics[width=0.49 \linewidth,clip]{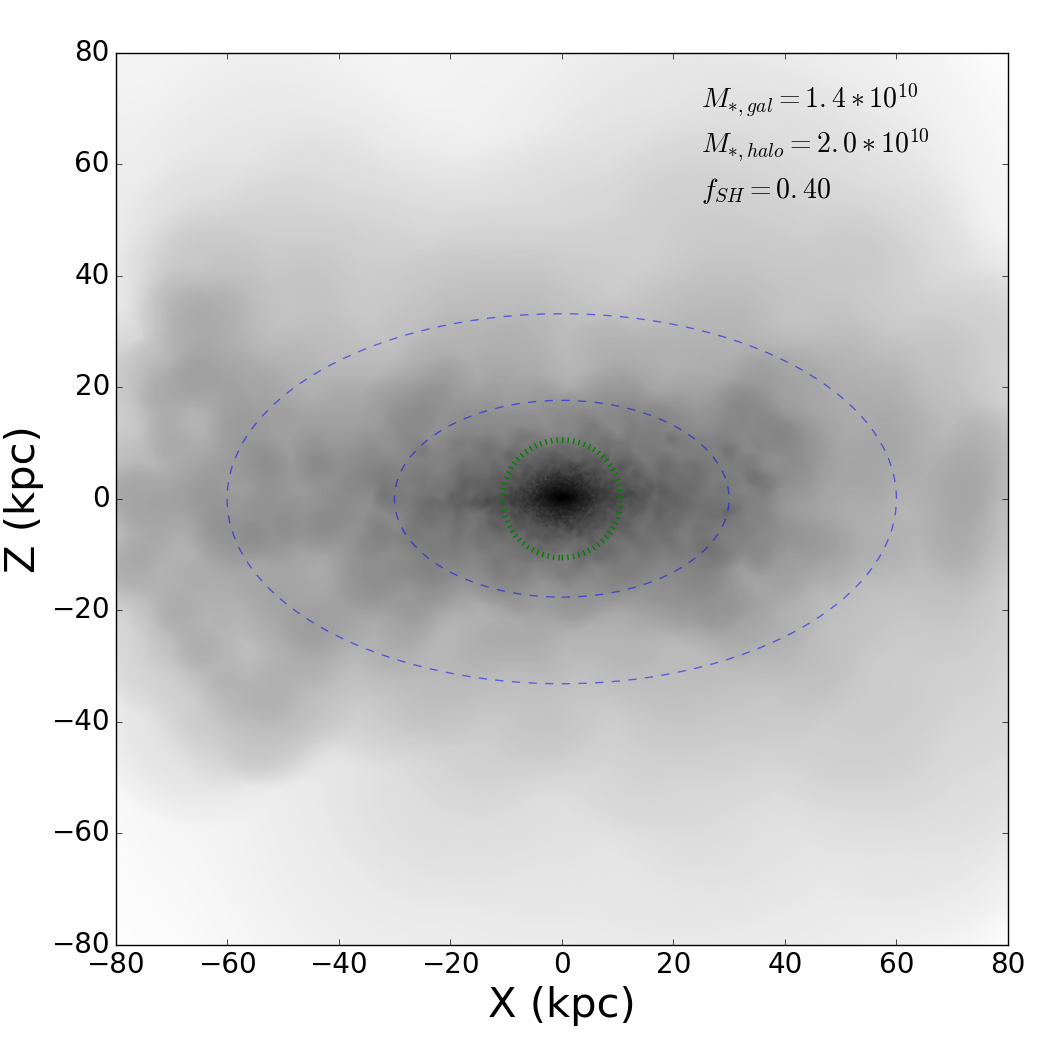}
\caption{Face-on (left) and edge-on (right) stellar projections of a
  randomly selected object in our sample. The inner green circle shows
  the separation between galaxy and stellar halo, defined here as all
  stars with distance smaller or larger than twice the half mass
  radius of the stars, $2r_h^*$, respectively. Dashed blue lines show
  the projected ellipsoids corresponding to the inertia tensor of
  stars in the stellar halo (i.e. $r >2r_h^*$) at two arbitrary
  radii $30$ and $60$ kpc. For comparison, the virial radius is $196$ kpc.
   The extended stellar halo in this object is
  preferentially oblated and aligned with the orientation of the inner
  galaxy, which is a common signature in our sample.}
\label{fig:snap}
\end{figure*}
\end{center}
%

Numerical simulations have shown remarkable successes at reproducing
several of the properties of stellar halos as well as at providing
insights into stellar halo formation \citep[e.g., ][]{Abadi2006,Abadi2003a,Abadi2003b,Helmi2008,Cooper2013,Tissera2014,Pillepich2015,Pillepich2014,Amorisco2017}. 
Recent simulations have shown that theoretical models also predict that the
stellar halo consists not only of a component accreted through mergers
but also of an in situ component made of stars that were formed
locally in the host galaxy.  The exact origin and fractions of in situ
stars in the stellar halo is not completely agreed upon. In situ stars
have been shown to dominate the inner regions of the stellar halo and
be displaced early in a galaxy's history by dynamical processes such
as mergers \citep{Pillepich2015,Zolotov2009,Tissera2014,McCarthy2012}. 
By contrast \citet{Cooper2015} find that the majority are formed as a result of the gas
stripped from infalling satellites. In situ stars have also been
claimed to be born in the stellar halo directly in localized
condensations of gas \citep{Tissera2013}. As different mechanisms are responsible for the
accreted and in situ components, one needs to address both in order to
understand: (i) how diversity in stellar halos arises and (ii)
specifically inspired by recent observations, how a
galaxy continues to assemble without growing its stellar halo.
A promising avenue to make progress on this question is to study the
emergent stellar halos in a
large statistical ensemble of galaxy assembly paths and merger histories
predicted by $\Lambda$CDM.

In this paper we take such an approach by using the large-volume
cosmological hydrodynamical simulation Illustris
(\textcolor{blue}{Vogelsberger et al. 2014a,Vogelsberger et
  al. 2014b,Genel et al. 2014}). Illustris 
is an ideal tool to tackle this problem
since it provides a sample of $1115$ central galaxies of mass
comparable to the Milky Way with predictions for both their dark
matter as well as baryonic assembly. This sample thus provides an
unbiased view of stellar halo formation, allowing further inspection
of dependencies on accretion history, environment and morphology of
the central galaxy among others. Although we sacrifice spatial
resolution compared to a zoom-in approach (Illustris' gravitational
softening length for stars is 0.7 kpc), the nature of the main quest
in this paper calls for large statistical samples and not for a
detailed view of a few individual galaxies, making the analysis of a
large cosmological box, such as Illustris, the most attractive
strategy. \nocite{Vogelsberger2014a,Vogelsberger2014b,Genel2014}

This paper is organized as follows: In Sec.~\ref{sec:sims} we briefly
describe the simulations and define our samples. We explore the
properties of stellar halos and their centrals in
Sec.~\ref{sec:diversity}, with specific emphasis in the populations
with the smallest and highest stellar halo fractions.  We turn our
attention to the in situ and accreted origin of stellar halos in
Sec.~\ref{sec:origin} and summarize our findings in
Sec.~\ref{sec:conc}.

\section{Numerical Simulations}
\label{sec:sims}

Illustris is a cosmological hydrodynamical simulation of a
representative volume of the Universe, comprising a box of size $106$ Mpc on a
side that was run with the moving grid code {\sc arepo}
\citep{Springel2010}.  In its high resolution version, Illustris-1,
the mass per particle is $1.6$ and $6.3$ $\times 10^{6} \; \rm M_\odot$
for baryons and dark matter, respectively, and a typical gravitational
softening length smaller than $0.7$ kpc \citep{Vogelsberger2014b}. The simulation is consistent with WMAP-9 standard cosmology, with $\Omega_m$=0.2726, $\Omega_b$=0.0456,
$\Omega_\Lambda$=0.7274, and $H_0$=70.4km/s/Mpc \citep{Hinshaw2013}. It
follows gravity by means of an octree algorithm and the
hydrodynamics of the gas including star formation and feedback
processes that are believed to be fundamental to the reproduction of the properties of
galaxies.

Gas cells can cool and heat self-consistently with a minimum
temperature floor of $T = 10^4 \rm K^\circ$
\citep{Vogelsberger2013}. Above a density threshold of $n_H=0.2 {\rm
  cm}^{-3}$, gas follows an equation of state that is used to
implicitly treat the multi-phase structure of the unresolved ISM
\citep{Springel2003}. Cold and dense gas above this threshold also
becomes eligible for star formation with a timescale proportional to
the local density. After stars form following a Chabrier initial mass
function they are evolved using Starburst99 \citep{Leitherer1999}
stellar evolution tracks. Stellar feedback is modeled kinematically,
by adding $100\%$ of the available energy due to supernovae as
momentum to surrounding gas cells. The mass loading of the wind scales
inversely with the local velocity dispersion of the dark matter
\citep{Vogelsberger2013}. Illustris also includes a treatment for
black hole growth and feedback, including a dual quasar/radio mode
\citep[see details in ][]{DiMatteo2005, Springel2005,Sijacki2015}, but
given the mass range and properties studied in this paper, our
conclusions are only minimally impacted by the physics of black holes.
The free parameters in the model were tuned to reproduce the global
star formation history of the universe and approximately follow the
observed stellar mass function at $z=0$. With those choices, Illustris
has been proven to successfully reproduce a series of {\it global}
properties of the galaxy population, such as scaling relations,
morphology bimodality, color distribution, merger rates, environmental
trends, among others
\citep{Genel2016,Rodriguez-Gomez2015,Sales2015,Sijacki2015,Snyder2015}.

Halos and galaxies are identified using the {\sc subfind} algorithm
\citep{Springel2001,Dolag2009}. In short, groups are first
identified based on a spatial-only information using the Friends of
Friends (FoF) algorithm and gravitationally self-bound
structures are later identified within groups by using {\sc
  subfind}. The object sitting at the center of the gravitational
potential of each halo is called the central or host galaxy and all other
substructures associated with the group will be referred to as
satellites or subhalos. We trace the history of each galaxy by means of the {\sc
  sublink} merger tree algorithm introduced in
\citet{Rodriguez-Gomez2015,Nelson2015} which allow the
user to trace the assembly history of any subhalo for up to 13.76 Gyr.

\begin{center}
\begin{figure}
\includegraphics[width=84mm]{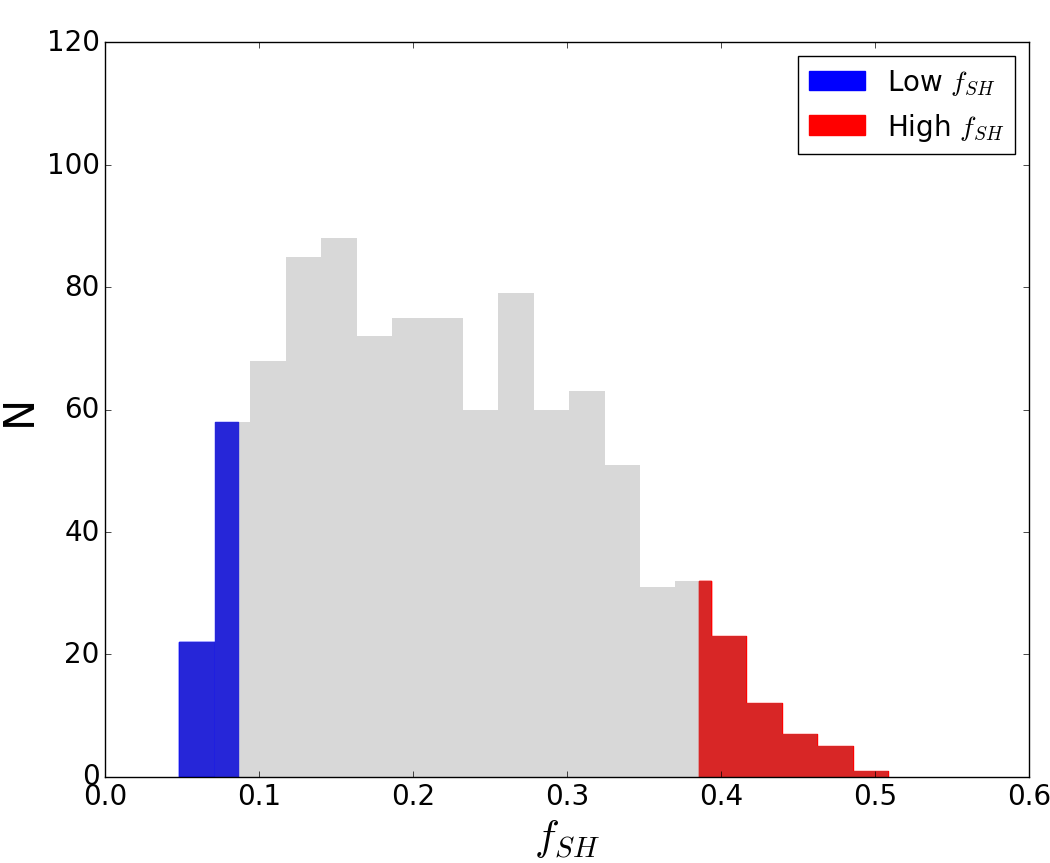}
\caption{Distribution of the stellar halo fraction, $f_{SH}=
  M_*(r>2r_h^*)/M_*^{\rm tot}$, of all central galaxies in
  Illustris within virial mass range
  $8.0 \times 10^{11} \; \rm M_\odot \leq M_{200} \leq 2.0 \times
  10^{12} \; \rm M_\odot$. There is
  a large variation in $f_{SH}$, with $5\%$ tails having low
  ($f<0.07$) and high ($f>0.45$) stellar halo fractions, highlighted
  by blue and red respectively. Each of these subsamples contains $60$
  galaxies.}
\label{fig:fhist}
\end{figure}
\end{center}

\subsection{Sample Selection}
\label{ssec:sample}

Our galaxies are selected to be comparable to the Milky Way by
imposing a virial mass \footnote{We define the virial mass, $M_{200}$, as that
  enclosed by a sphere of mean density $200$ times the critical
  density of the Universe, $\rho_{\rm crit}=3H^2/8\pi G$.  Virial
  quantities are defined at that radius, and are identified by a
  ``200'' subscript.} cut $M_{200}=8 \times 10^{11}$-$2\times10^{12}
\; \rm M_\odot$ and selecting central galaxies only (i.e. no
satellites). A total of 1115 central galaxies lie in this range. We define as a
``galaxy'' all stars or gas elements that are within a galactic radius
defined as $r_{\rm gal} = 2 r_h^*$, where $r_h^*$ is the half-mass radius of the
stars as computed by {\sc subfind}. Each central galaxy has stellar
particles associated to it (and gravitationally bound) all the way out to
the virial radius. Stars that are located beyond $r_{\rm gal}$ that do
not belong to any satellite and that are within $r_{200}$ will be
assigned to the stellar halo component.

We define stellar halo fraction as:
\begin{equation}
f_{SH} = \frac{M_{*,\rm tot}-M_*}{M_*}
\end{equation}
where $M_{*,\rm tot}$ is the total stellar mass of the halo and $M_*$
is the stellar mass of the halo within $r_{\rm gal}$. Although there
is no consistent definition of stellar halo across the field, similar
definitions have been adopted in previous studies
\citep[e.g. ][]{Pillepich2015,Merritt2016}.

Fig.~\ref{fig:snap} shows an example of one of our simulated galaxies and
its halo. Using the angular momentum vector of the central galaxy
(i.e. stars within $r_{\rm gal}$) we define face-on (left) vs. edge-on
(right) projections. Stars distribute well beyond the galaxy radius 
$r_{\rm gal}$ shown by the green dotted circle,
forming a diffuse and extended stellar halo reaching out beyond $100$
kpc. Stellar halos are often inhomogeneous
with the presence of substructure indicating the occurrence of
relatively recent merger events in the past. Note that these halos are
far from spherical, as indicated by the blue ellipsoids showing the
inertia tensor axis-ratios for stars at two different radii
(calculated with all stars outside $r_{\rm gal}$ but within the radius
of interest), a subject we return to in Sec.~\ref{sec:shapes}.

\begin{center}
\begin{figure}
\includegraphics[width=84mm]{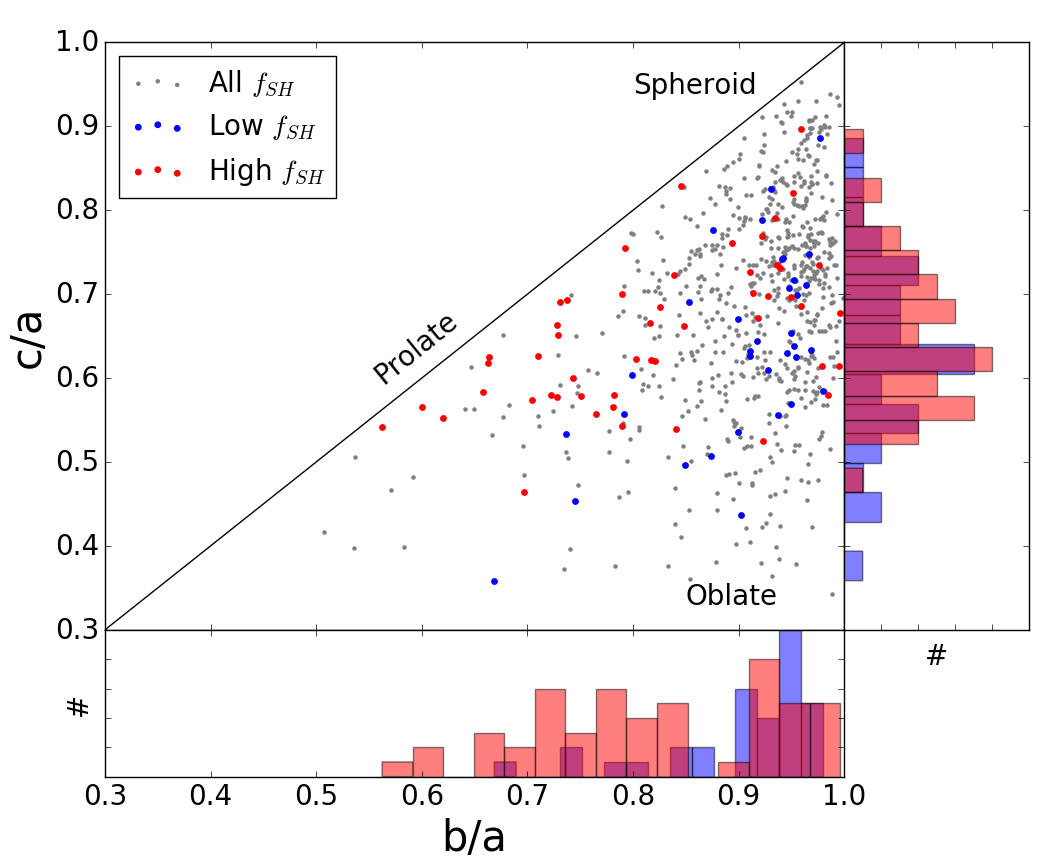}
\caption{Stellar halo shapes in the Illustris simulation for galaxies
  in the virial mass range $8 \times 10^{11} < M_{200}/ \rm M_\odot <
2\times 10^{12}$. $a$, $b$ and
  $c$ correspond to the major, intermediate and minor axis lengths of the 
  ellipsoid with inertia tensor defined by stars in the stellar halos. Grey shows
  the whole sample and highlighted in blue and red are the low and
  high stellar halo fraction samples, respectively, as defined in
  Fig.~\ref{fig:fhist}. The distribution favors oblate halos overall,
  with no particular dependence on the fraction of stellar
  halo.}
\label{fig:shape}
\end{figure}
\end{center}
%
%
We exclude centrals that have massive
satellites ($M_{*}\geq 10^9 \; \rm M_\odot$) within $50$ kpc from the
center as well as centrals with a satellite of stellar mass
$M_{*,sat}\geq \frac{1}{4} M_{*,cen}$ anywhere within the virial
radius ($r_{200}$). These two conditions ensure that the stellar halo
mass estimates will not be enhanced artificially by stars in the
outskirts of satellites that are mistakenly assigned to the central by
limitations of our subhalo finder. With these definitions, our sample comprises 
of 967 centrals and spans a wide range in stellar
halo fraction, $0.05 \leq f_{SH} \leq 0.51$, shown in
Fig.~\ref{fig:fhist}.  Understanding the origin of such diversity is
one of the main objectives of this paper. To that end, we will compare
the properties and evolution of both tails of this distribution, 
which we will refer to as the ``low/high stellar halo fraction'' samples
and use blue and red respectively to denote them throughout this work. The low
and high $f_{\rm SH}$ samples are selected to be the $5\%$ tails, with a
total of $60$ galaxies each.  
\section{Shape and surface brightness profiles of stellar halos}
\label{sec:shapes}

We start by quantifying the shapes of the simulated stellar halos. 
 Fig.~\ref{fig:shape} shows the $b/a$ and $c/a$ axis
ratios, where $a$, $b$ and $c$ are the square roots of the eigenvalues corresponding to
the major, intermediate and minor axes of the inertia tensor of the
stars in each stellar halo of our sample. We adopt the definition from
\citet{Vera-Ciro2011}, for which the normalized inertia tensor is:

\begin{equation}
I_{ij} = \sum_{x_k \in V} \frac{x_k^{(i)} x_k^{(j)}}{d_{k}^2}
\end{equation}

\noindent
where $d_k$ is a distance measure to the kth star and V is the set of
stars in the stellar halo. We use all stars in the stellar halo to
assign a single $b/a$ and $c/a$ to each of our
centrals. Fig.~\ref{fig:shape} shows that the majority of stellar
halos are not spherical and, in fact, they occupy almost all available
space in the axis ratios map with a slight tendency for oblated
distributions (large $c/a$). This trend is stronger for galaxies with
low stellar halo component than with large $f_{\rm SH}$ and is in good
agreement with observational estimates of halo shapes in disky
galaxies \citep{Harmsen2017}.  Color coding follows from
Fig.~\ref{fig:fhist}, with gray for the whole sample of MW-like
objects and blue or red highlighting our low and high $f_{\rm SH}$
subsamples respectively.  We have checked that for the cases where the
central galaxy has a well-defined disk, the shapes of the stellar
halos tend to align with the inner disk, as is the case in the example
shown in Fig.~\ref{fig:snap}.

\begin{center}
\begin{figure}
\includegraphics[width=84mm]{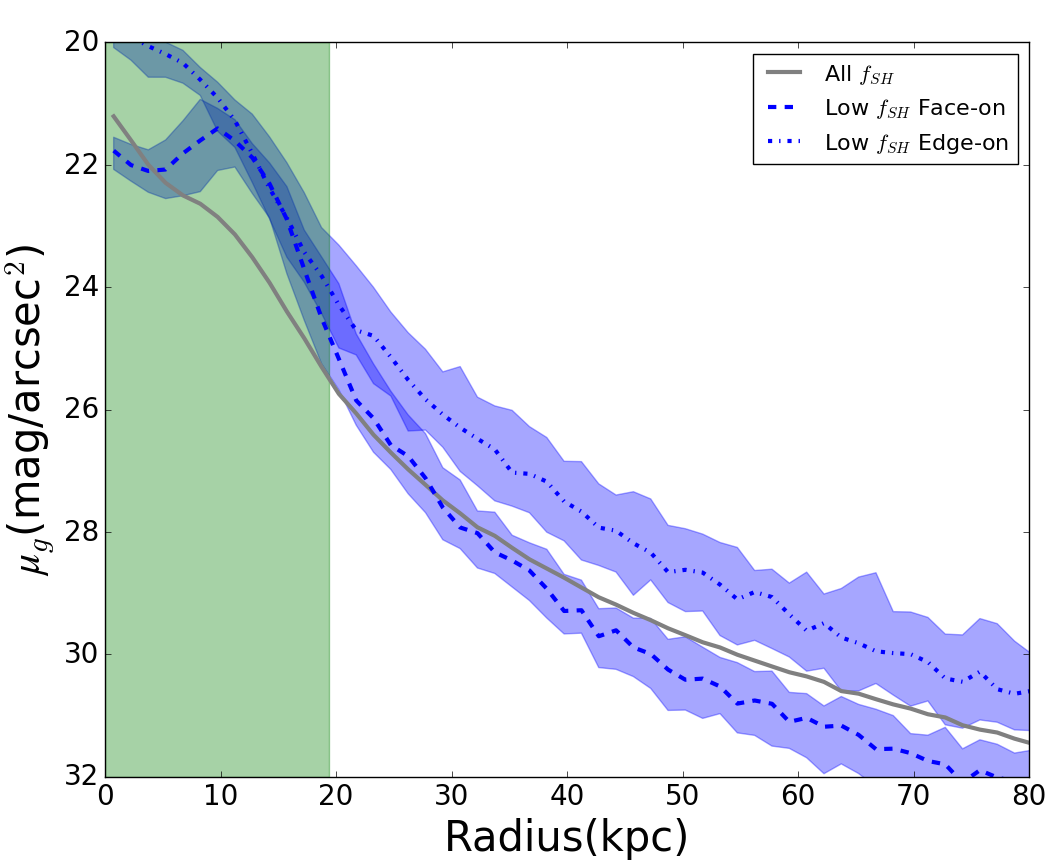}
\caption{Projection effects on the surface brightness profiles of
  stellar halos. There is a sizable effect according to the
  orientation of galaxies, with face-on projections (dashed line)
  giving a fainter stellar halos outskirts than when the galaxy is
  edge-on (dotted line). The effect can be as large as $2$ magnitudes
  fainter in the outskirts of the halos.}
\label{fig:proj_effect}
\end{figure}
\end{center}

This asphericity can introduce observational biases in stellar halo
studies, in particular for those which targets are selected based on
inclination of the disk. Consider, for instance, an oblated
distribution, $a \sim b \gg c$. If we ``observe'' the object face-on,
the projection will be done along the short axis rendering a
considerably lower stellar density than if the object was observed
edge-on where the projection is done along the more extended axes $a$
or $b$. With the typical axis ratios found in our simulations, the
effect can be sizable, as shown in Fig.~\ref{fig:proj_effect}. Thick
dotted and dashed lines show the median projected mass surface density
profile of our sample with low $f_{\rm SH}$ when the galaxies are all
oriented face-on or edge-on, respectively, based on the angular
momentum of the stars within $r_{\rm gal}$.  The profiles are computed
in integrated ellipsoidal shells, with the horizontal axis showing the
ellipsoidal radius as computed from the inertia tensor axis ratios
($b/a$ or $c/a$ depending on the projection). This procedure is
equivalent to the common practice in observations of measuring the
axis ratios of the isodensity contours on light profiles and scaling
the ellipsoidal radius accordingly.

\begin{center}
\begin{figure*}
\includegraphics[width=0.489\linewidth]{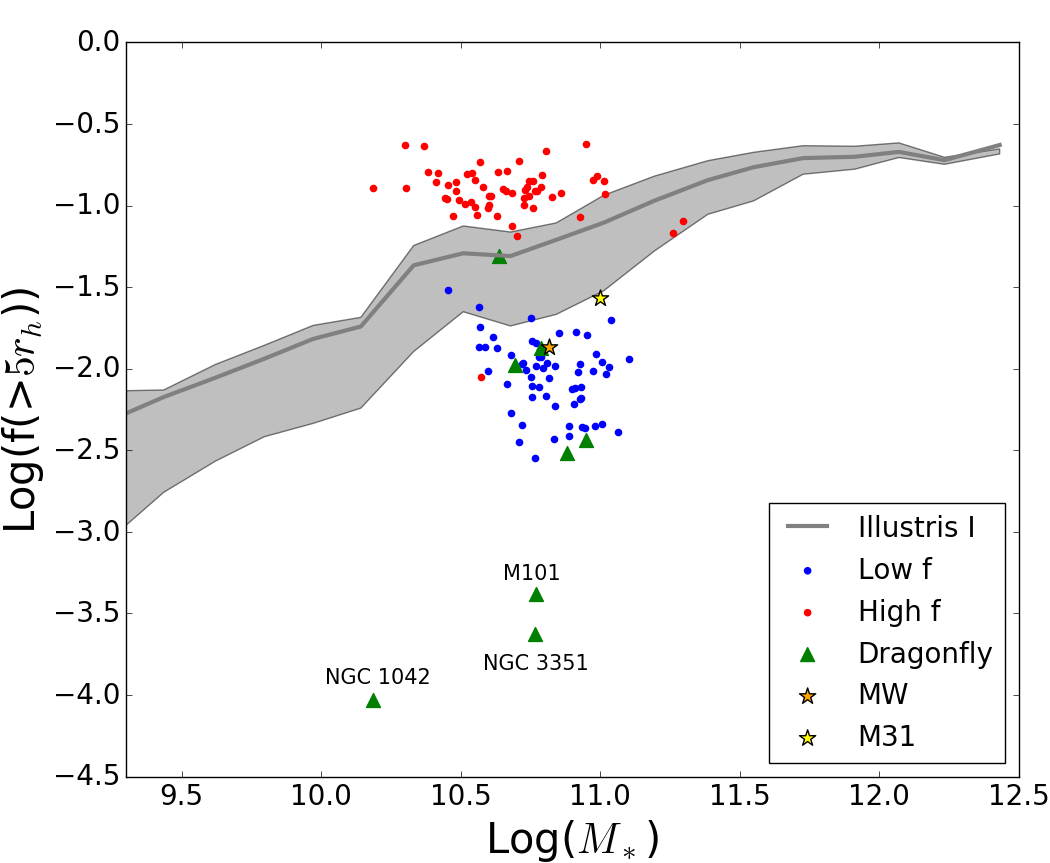}
\includegraphics[width=0.489\linewidth]{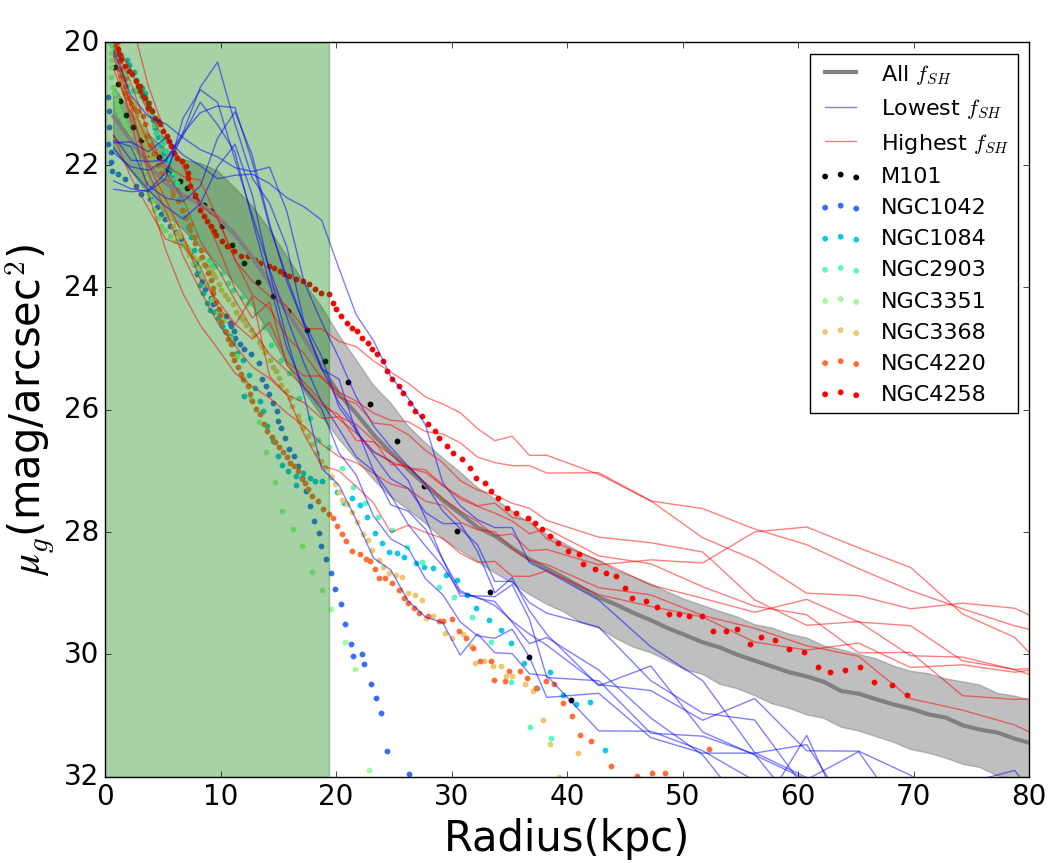}
\caption{{\it Left}: Stellar halo fraction calculated as the mass
  beyond $5*r_h^*$ for easy comparison to observations. Color coding as
  before shows in gray the simulated sample and highlighted red and
  blue points correspond to the high and low $f_{SH}$ galaxies,
  respectively. The sample overlaps well with data from the Dragonfly
  Telephoto Array, shown in green triangles \citep{Merritt2016}. M31 (yellow)
  and the MW (orange) are also indicated with a starred symbol
  \citep{Carollo2010,Courteau2011}. A few objects stand out from the observational sample
  having a very low and almost undetected stellar halo, such as M101,
  NGC3351 or NGC1042. {\it Right}: Median (face-on) surface brightness profile
  for the simulated stellar halos is shown in gray solid line with
  shading indicating the $25\%$-$75\%$ percentiles in our sample. A
  green vertical region is used to indicate twice the average half-mass
  stellar radius of the stars, and therefore the division between the
  average  ``galaxy'' regime, and the stellar
  halo extending beyond that. 
  For comparison, data from disk galaxies in \citet{Merritt2016} is
  shown with colored dots. Thin blue/red lines indicate individual profiles
of the 8 most extreme objects in Illustris with low and high $f_{\rm SH}$ respectively.}
\end{figure*}
\label{fig:obs}
\end{center}
%

Our tests indicate that solely changing the orientation of the
projection for the {\it same} simulated objects can result in a
difference of up to $2$ $g$-band magnitudes. This effect will
artificially introduce variations among observed stellar halo profiles
and should be taken into account when analyzing stellar halo diversity
in real galaxies. {\it One prediction from our study is that surface
  brightness profiles of face-on oriented disks will be biased low
  with respect to a more inclined sample}. Interestingly M101, a
galaxy put forward by several authors as having little to no stellar
halo, has an almost perfect face-on projection (inclination angle
$i=16.0$, from the HyperLEDA database \citep{Makarov2014}). The other
two cases with no stellar halos drawn from the Dragonfly study in
\citet{Merritt2016}, NGC1042 and NGC3351, are less clear. With
inferred inclinations of $i=58.1$ and $54.6$, they would show a milder
effect due to projections compared to M101. A more conclusive analysis
will come once more observations of galaxies and their faint stellar
halos become available. For now, we just highlight the need for
addressing projection effects due to stellar halo shapes and, in
particular, the derived conclusions on stellar halo mass and
structure.

\begin{center}
\begin{figure}
\begin{subfigure}{.5\textwidth}
\includegraphics[width=0.89\linewidth]{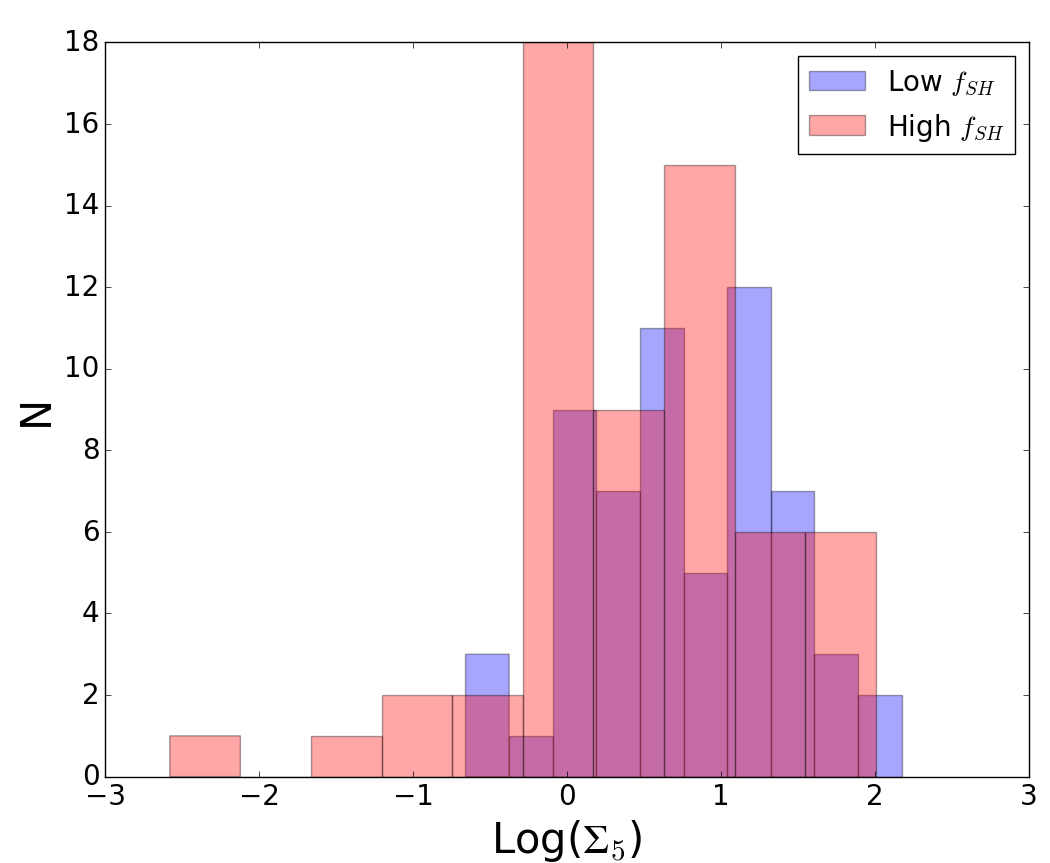}
\end{subfigure}
\begin{subfigure}{.5\textwidth}
\includegraphics[width=0.89\linewidth]{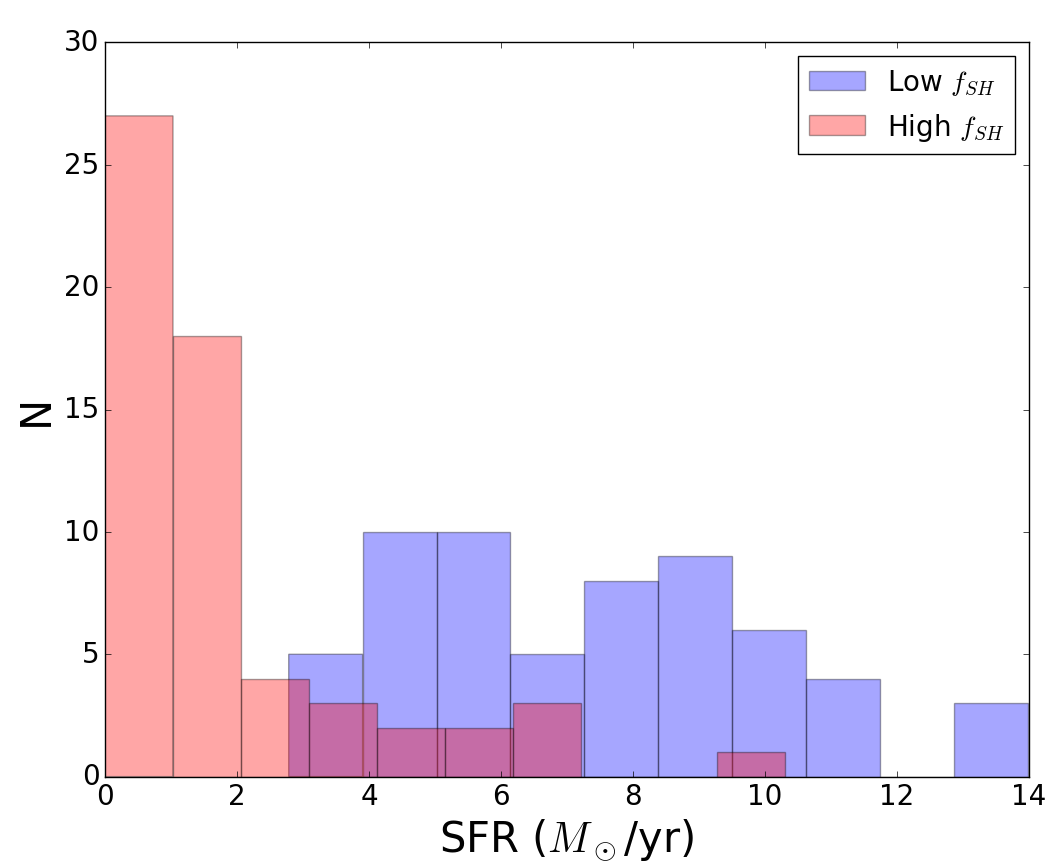}
\end{subfigure}
\caption{Distribution of environment (as defined by the
  distance to the fifth neighbor $\Sigma_5$) in the top and the star
  formation rate (SFR) in the bottom panels for the low and high
  stellar halo fraction samples. Whereas there is no obvious trend
  with environment, there is a clear difference in the SFR between
  both samples, with low $f_{\rm SH}$ galaxies being more actively star
  forming than the galaxies with the highest stellar halo content.}
\label{fig:props}
\end{figure}
\end{center}

With these caveats in mind, we compare our simulated samples to
available observations in Fig.~\ref{fig:obs}. On the left panel we
show the stellar halo fraction as a function of stellar mass of the
central galaxy. To more closely match the definition introduced in
\citet{Merritt2016}, we consider the fraction of the stellar mass that
is beyond $5r_h^*$ instead of the $2 r_h^*$ adopted for the rest of
our paper. For simplicity, this quantity in simulations is computed in
3D instead of in projection as is done in observations.  The gray
solid line shows the median for a representative number of central
galaxies on a wide mass range in Illustris, with the shaded area
corresponding to the 25th to 75th percentiles. Red and blue dots
indicate the values for our low and high stellar halo fraction
galaxies that according to our selection criteria cluster around
MW-like objects, $M_*\sim 6 \times 10^{10}\; \rm M_\odot$.

As expected, the blue and red subsamples represent the tails of the
normal population, even though the selection was made using $2r_h^*$
instead of $5$. We note that there is one central from the high $f_{
  \rm SH}$ sample with a particularly low value of $f_{5rh}$. This
galaxy is undergoing a merger that is too minor to qualify for our
conditions, but is affecting the subhalo finder.  Results for real
observed galaxies are shown in green triangles (Dragonfly) and
yellow/orange starred symbols for the Milky Way and M31,
respectively. Although no extreme outliers such as M101, NGC3351 and
NGC1042 seem to appear in Illustris, the simulated sample occupies
roughly the same region as the observed stellar halos in
literature. The good agreement is particularly encouraging if one
takes into account the difference in definitions (3D vs. 2D) and
procedures (i.e. background subtraction) that go into computing such
fractions. Simulations with low $f_{\rm SH}$ (blue) seem to better
represent the observations in the mass range of overlap.  As we will
see in Sec.~\ref{ssec:hosts}, this can be easily explained as a
morphological bias in observed samples as we are considering all
simulated galaxies in the mass range with no morphological cuts.

The right panel of Fig.~\ref{fig:obs} confirms that the median surface
brightness profiles of simulated galaxies (shown in gray) is also in
reasonable agreement with individual observations from Dragonfly
(colored symbols), although there is a trend in Illustris to have
slightly more massive stellar halos overall. In particular, we are
unable to find good analogs to the lowest stellar halo mass objects
such as M101 in Illustris. The averaged more massive stellar halos
in the simulations is already hinted on the left panel of Fig.~\ref{fig:obs}
and is mainly driven by the fact that observations typically target disk-dominated
galaxies whereas we are including all galaxies in a halo mass range,
regardless of their morphology. In addition to this, one cannot
disregard the possibility that stellar halos in Illustris are
overpredicted due to the over abundance of low mass galaxies in the
simulation (see top right panel in Fig.2 of \citealt{Genel2014}), which
presumably can also be tidally disrupted, contributing more stars than
they should to the stellar halos. This, however, does not represent a
major limitation of our analysis as the main objective is to
understand the {\it variations} on the stellar halo fractions in a
narrow range of halo mass, and not to exactly match individual
galaxies from observations. Indeed, as highlighted in
Fig.~\ref{fig:fhist} and \ref{fig:obs}, the diversity shown in $f_{\rm
  SH}$ for our simulations is significant and comparable to
observations, making Illustris and its large statistical sample a
powerful tool to study the emergence of diversity in stellar halos
within $\Lambda$CDM.

\section{Building the Stellar Halos' diversity}
\label{sec:diversity}

\subsection{Properties of the host galaxy}
\label{ssec:hosts}

Our low/high $f_{\rm SH}$ samples are selected purely on the fraction
of mass in their stellar halos compared to that of the stars in the
host galaxy which is essentially blind to the properties of the
central galaxy.  It is therefore interesting to explore how the
central galaxies in each $f_{\rm SH}$ subsample compare to each
other. We start by looking at the environment of galaxies with the
lowest and highest stellar halo fractions in the top panel of
Fig.~\ref{fig:props}. Environment has been quantified by means of the
mean stellar density $\Sigma_5$, defined as the stellar mass density
enclosed within the distance to the $5th$ nearest neighbour with
stellar mass above $M_*=10^{10}$. In our sample, the average distance to
the $5th$ neighbour is $r_5 = 4.8$ Mpc.  We find no significant
difference between the low and high $f_{\rm SH}$ which suggests that
the medium-scale vicinity of a halo at present day encodes little
information on the formation of stellar halos. This lack of
correlation with environment is in agreement with trends found in
observations \citep[see e.g. ][]{Merritt2016}.
%
\begin{center}
\begin{figure}
\includegraphics[width=84mm]{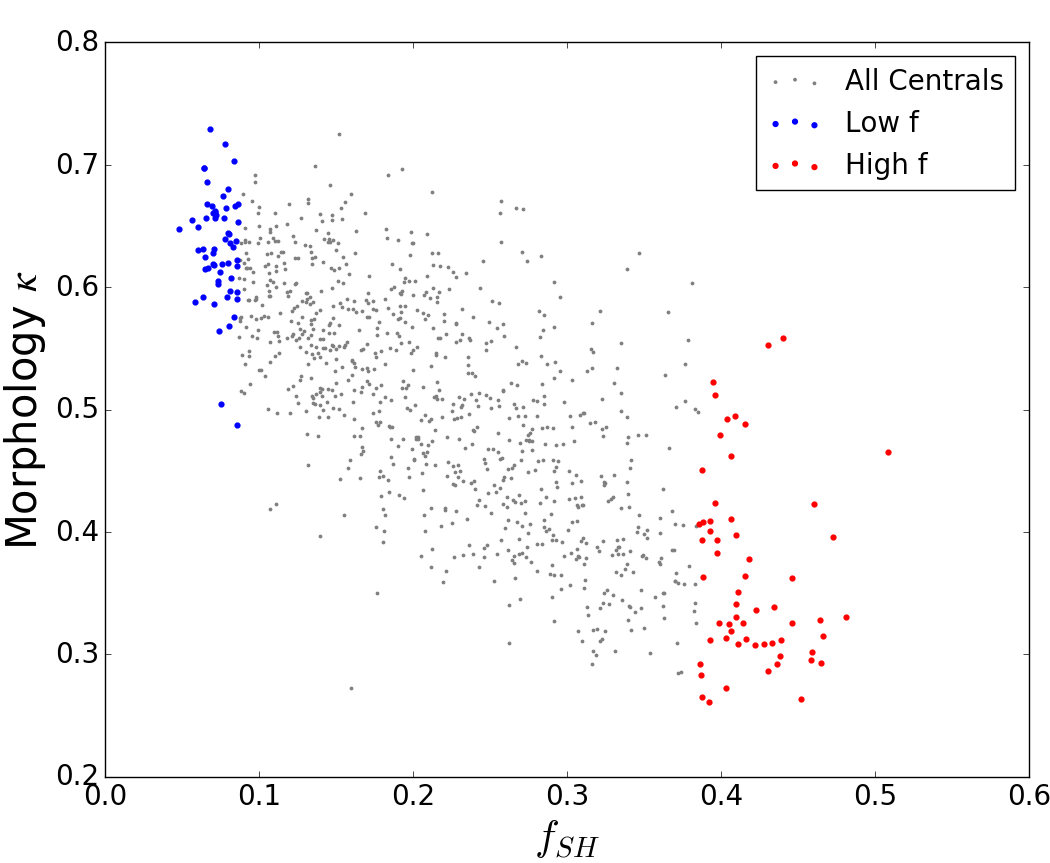}
\caption{There is a well-defined correlation between the fraction of
  stellar halo $f_{\rm SH}$ and galaxy morphology as quantified
  kinematically through $\kappa$.  Disk-dominated galaxies (large
  $\kappa$ values) show the least amount of stellar halo fractions
  with average $f_{\rm SH} \sim 0.15$. The trend of $f_{\rm SH}$ with
  morphology is much stronger than the relevance of accreted stars or
  mergers {\it within} the galaxy radius \citep[see e.g.,
  ][]{Sales2012,Rodriguez-Gomez2016}. }
\label{fig:kappa}
\end{figure}
\end{center}
%
On the other hand, there is a clear difference in the star formation
activity of centrals according to their stellar halo fraction (see
bottom panel Fig.~\ref{fig:props}), with high stellar halo content
centrals being preferentially non-star forming in contrast with
galaxies with low $f_{\rm SH}$ which are actively forming stars. {\it
  This correlation is also present in the morphology of the galaxies,
  with disk-dominated galaxies having on average a significantly lower
  fraction of stars in the stellar halo compared to the central galaxy
  than early-types}. This can be seen in Fig.~\ref{fig:kappa}, where
morphology $\kappa$ is estimated following the methodology of
\citet{Sales2012}, as the ratio between the kinetic energy of stars in
ordered motion around the $z$-axis and the total kinetic energy of the
system. Large $\kappa$ values correspond to disk-dominated morphology,
whereas a low $\kappa$ index is indicative of spheroidal morphologies
with no dominant rotation. Although the scatter is significant, the
correlation between $f_{\rm SH}$ and $\kappa$ is well defined in our
sample and agrees well with the observational trends reported in
stacked SDSS observations of stellar halos for galaxies of different
morphological types \citep{DSouza2014}.

The link between stellar halo content and morphology, although not
completely unexpected, is still nontrivial. For MW-mass objects as
analyzed here, mergers have been shown to correlate very poorly with
the central morphology of the galaxy
\citep{Sales2012,Rodriguez-Gomez2017}, yet they are believed to build
the majority of the stellar halo. How does this correlation then get
established? From Fig.~\ref{fig:kappa} it is clear that the processes
responsible for shaping the morphology of galaxies also have a large
impact on the formation of the stellar halo. We turn our attention to
the general assembly of the galaxies and their dark matter halos for
further clues.

\subsection{Assembly histories for dark and stellar halos}
\label{ssec:merg}

\begin{center}
\begin{figure}
\includegraphics[width=84mm]{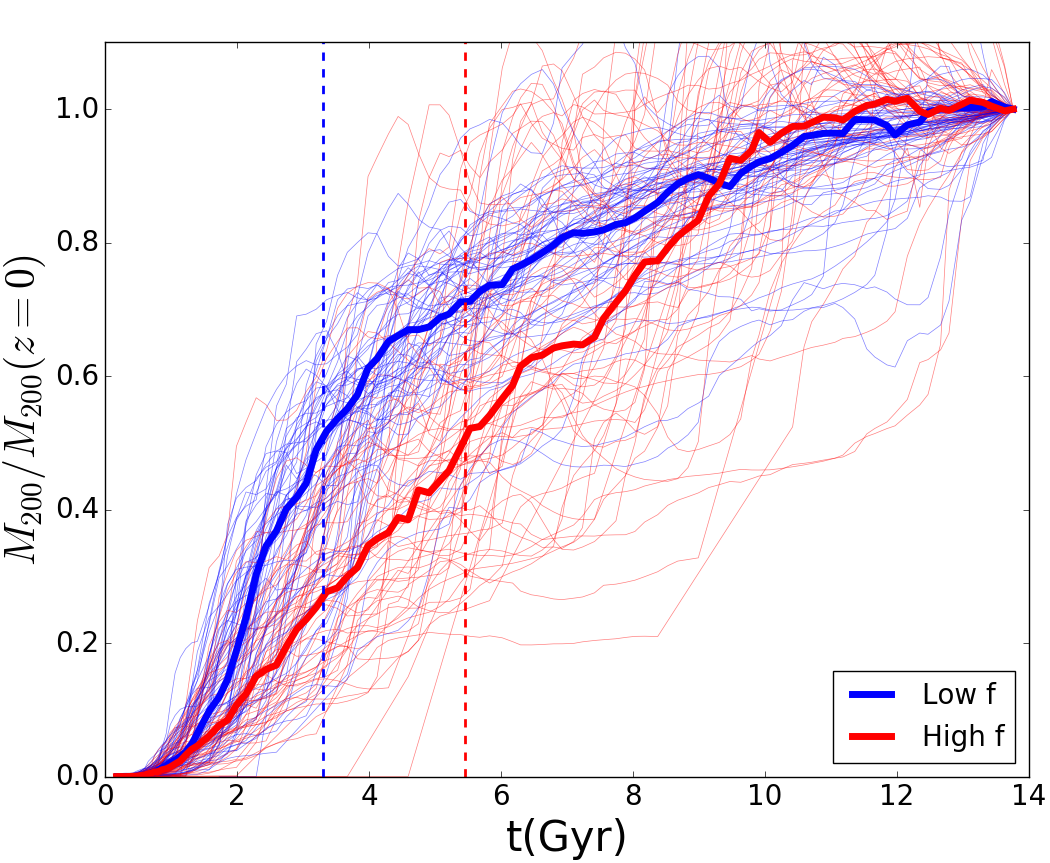}
\caption{Virial mass accretion history as a function of time for
  galaxies with low (blue) and high (red) stellar halo fraction.
  Thick lines indicate the median and thin lines show individual
  tracks for each halo. Galaxies with low $f_{\rm SH}$ show an earlier
  total mass assembly, reaching half their present-day mass by $t\sim
  3$ Gyr compared to $t \sim 5.5$ Gyr for the objects in the high
  stellar halo fraction subsample. The late-time evoution in the last
  $3$ Gyr is similar for both groups.}
\label{fig:mvst}
\end{figure}
\end{center}

Fig.~\ref{fig:mvst} shows the virial mass $M_{200}$ assembly histories
for our objects in low (blue) and high (red) $f_{\rm SH}$
subsamples. To ease the comparison, the vertical axis shows the virial
mass at a given time normalized to the final mass at $z=0$ such that
most curves stay within the $0$-to-$1$ range.  Thin lines indicate
tracks for each individual object and the thick curve shows the median
of each subsample. Although the object-to-object scatter is large, the
median trend suggests a different mass assembly history characteristic
for each subsample.  Galaxies with low stellar halo fractions assemble
their halo mass earlier compared to objects with larger $f_{\rm SH}$,
as shown by the blue and red thick lines. For instance, galaxies with
low $f_{\rm SH}$ have acquired on average half of their mass by
$t_{50}=3$ Gyr ($z_{50}=2.2$) whereas the counterparts with large
stellar halo fraction take double the time to reach that value, or
$t_{50} \sim 6$ Gyr. 

Moreover, the overall shapes of the assembly histories are different.
High $f_{\rm SH}$ objects show a
steady growth throughout most of their history, slowing down slightly
after $t \sim 10$ Gyr. Instead, in galaxies with low $f_{\rm SH}$ the
growth is much more rapid in the initial phase ($t \leq 4$ Gyr), gradually
slowing down afterwards. Interestingly, the late time
evolution of $M_{200}$ for both subsamples is comparable after $t \sim
9$ Gyr, or equivalently, $z=0.45$. It is worth highlighting that the
median virial masses of the low and high $f_{\rm SH}$ subpopulations
are $1.1$ and $1.4  \times 10^{12} \; \rm M_\odot$ respectively, which
partially explains the earlier collapse of galaxies with low stellar
halo fractions (as lower mass halos collapse earlier). However, the
bias in halo mass is not large enough to fully explain the
difference in median formation history found in Fig.~\ref{fig:mvst}.

\begin{center}
\begin{figure}
\includegraphics[width=84mm,height=100mm]{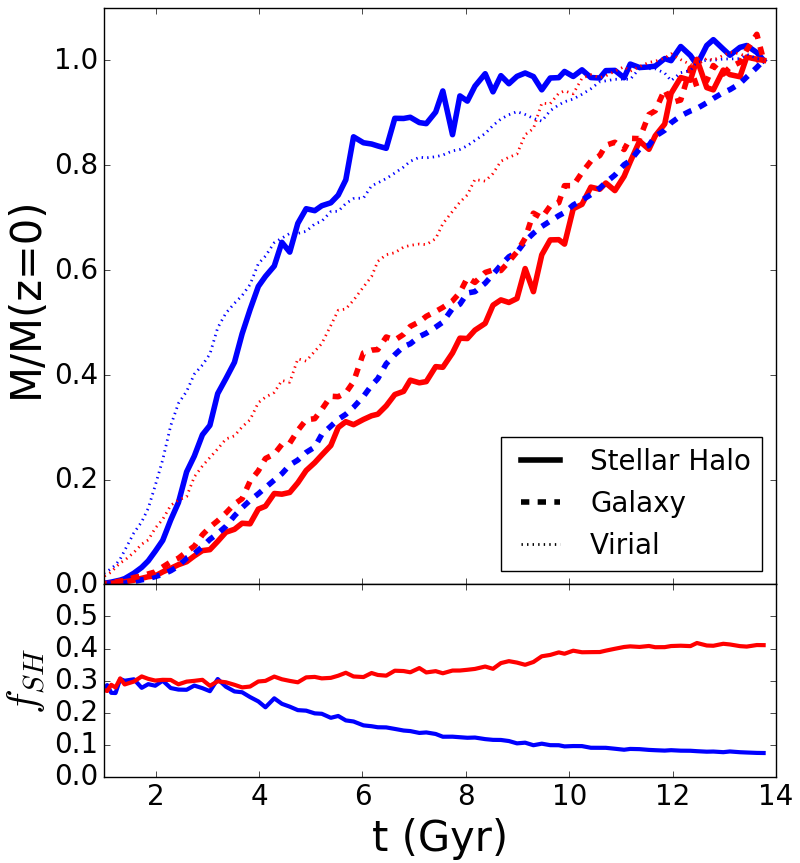}
\caption{{\it Top}: similar to Fig.~\ref{fig:mvst}, but for the
  assembly history of the {\it stellar halo} mass. Thick solid lines
  show the median for the low and high $f_{\rm SH}$ sample and
  confirms that the stellar halos of galaxies with low $f_{\rm SH}$
  also assemble earlier on average compared to those stellar halos in
  the high $f_{\rm SH}$ sample, in agreement with the trend found for
  virial mass (shown in thin dotted curves). In contrast, the mass
  assembly of the central galaxies (long-dashed curves) show no
  statistical difference between both samples. This then means that
  the evolution of the stellar halo fraction $f_{\rm SH}$ with time
  (bottom panel) is similar in both samples at the beginning ($t \leq
  3$ Gyr), but as the low $f_{\rm SH}$ stops growing their halos but
  continue to grow their central galaxy, their overall stellar halo
  fraction declines with time afterwards. On the other hand, the red
  curve grows their stellar halo faster than their central mass,
  causing them to populate the largest $f_{\rm SH}$ of the population
  at $z=0$. }
\label{fig:mvst_all}
\end{figure}
\end{center}

If the dark matter halos of galaxies with low stellar halo assemble
earlier, one is left to wonder whether the same applies to the stellar
assembly. We explore this in Fig.~\ref{fig:mvst_all}, where solid
lines correspond to the median assembly of the stellar halo component
and dashed curves indicate the stellar assembly of the central
galaxy. To guide the eye, we have included the median assembly of
$M_{200}$ from the previous figure in thin dotted lines. To calculate
these curves, we trace the progenitors of our galaxies backwards in
time and at each snapshot count the amount of mass within (for $M_*$)
and beyond (for the stellar halo mass $M_{SH}$) the instantaneous
galaxy radius $r_{\rm gal}$ defined as before, $r_{\rm gal} =
2r_{h,*}$, with $r_{h,*}$ the half mass radius of the stars in the
galaxy at each time. We then take the median mass at a given time from
each subsample. We highlight that these curves measure the mass
assembly onto each object and not the formation time of the stars.

The trends in Fig.~\ref{fig:mvst_all} are intriguing and suggest that
the assembly of the mass in the stellar halo component itself is
different between both samples. Galaxies with low $f_{\rm SH}$ build
their stellar halo early on in a similar fashion as the virial halo
build up: a rapid early growth followed by a slow down of the
evolution at later times. For instance, $80\%$ of the stars in the
stellar halos are already in place at $t=6$ Gyr while later evolution
during half a Hubble time only adds the remaining $20\%$. Note that
this behaviour is different from the stellar mass growth proceeding 
in the central galaxy, which is almost constant with time
(see blue dashed curve), with a median time of $12$ Gyr to build
$80\%$ of the stars in the center. On the other hand, galaxies with large
stellar halo content assemble both the mass in their stellar halos {\it
  and} in the central galaxy at a steady rate which is only loosely
connected to the $M_{200}$ growth.
%

The combined effect of stellar halo and central stellar growth is
shown in the bottom panel of Fig.~\ref{fig:mvst_all}. By construction,
both our subsamples have very different $f_{\rm SH}$ today. However,
this seems not to have been always the case. The early progenitors of
both samples had similar fractions of mass in the stellar halo compared
to the central until $t \sim 4$ Gyr, after which both subsamples start
to deviate, an effect governed mainly by the fact that the low $f_{\rm
  SH}$ sample has significantly slowed down the building of its
stellar halo but continues to actively grow the stellar mass in the
central object bringing the $f_{\rm SH}$ ratio down with
time. {\it Galaxies with low stellar halo content might be the 
most ancient stellar halos ever assembled}.

\begin{center}
\begin{figure*}
\includegraphics[width=0.485\linewidth]{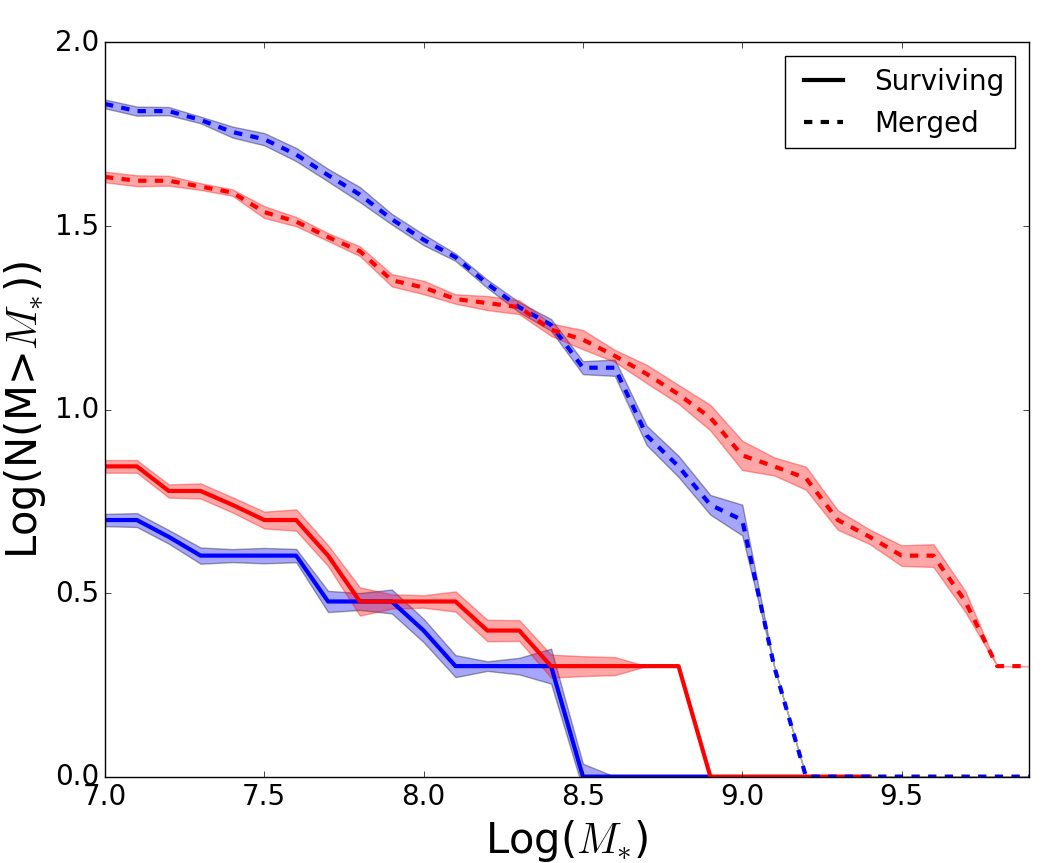}
\includegraphics[width=0.485\linewidth]{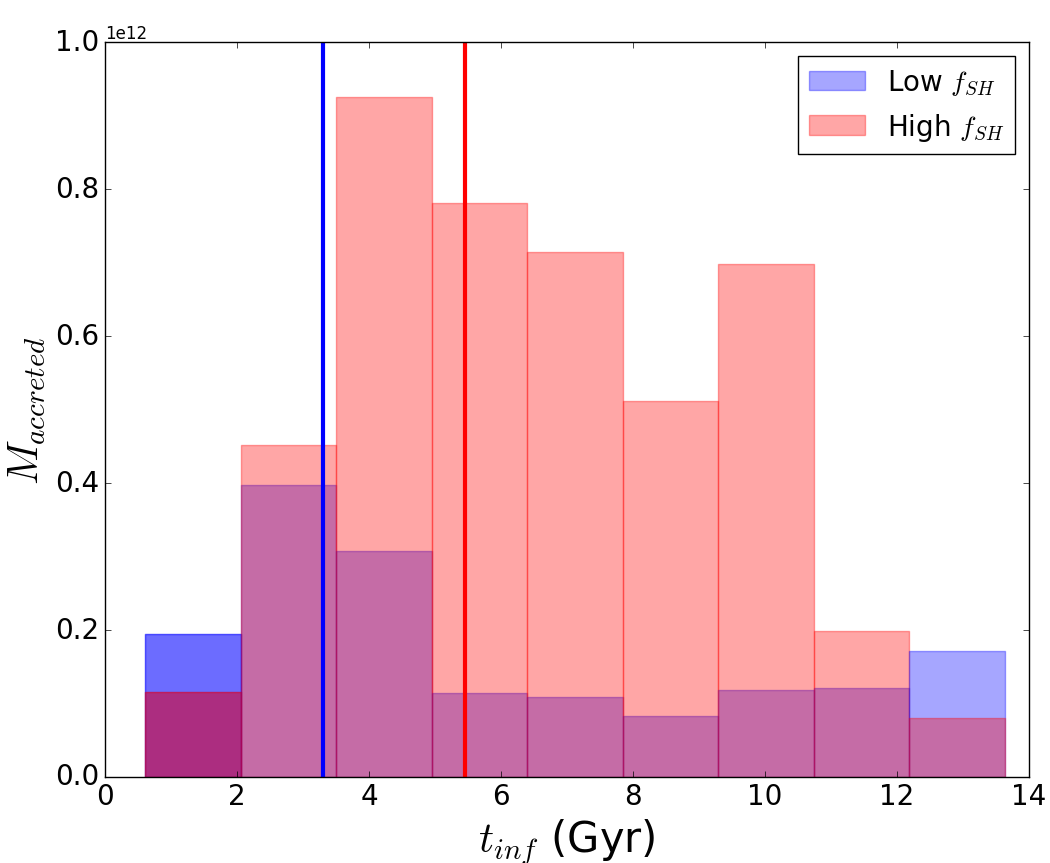}
\caption{{\it Left:} Stellar mass functions for merged (dashed) and
  surviving (solid) satellites. Red and blue lines are medians of the
  high and low $f_{SH}$ populations, respectively. High $f_{SH}$
  galaxies have, on average, both more merged and more surviving
  satellites than low $f_{SH}$ population, but the effect is small
  ($\sim 30\%$) and independent of satellite mass.  {\it Right:}
  Histogram of infall times weighted by stellar mass for merged
  satellites. The infall time, $t_{\rm inf}$, is defined as the time
  at which a satellite reaches its peak total (dark matter, gas and
  stars) mass. The accreted stellar mass for the low $f_{\rm SH}$
  population infalls earlier compared to that of the high stellar mass
  fraction. For comparison, the time for the virial mass $t_{50}$ is
  shown and roughly reproduces the peak time of infall for disrupted
  satellites.}
\label{fig:satmass}
\end{figure*}
\end{center}

An important ingredient in the evolution of stellar halos is the
population of satellite galaxies, in particular, those satellites that
get partially or totally tidally disrupted contributing their stars to
the extended halo\footnote{The reader should bear in mind that
  disrupted satellites can contribute stars to the stellar halo or
  central galaxy, depending on the position where they distribute
  their stars}. Surviving satellites do also donate some of their
stars to the stellar halo, and their contribution to the stellar halo
can vary from negligible to significant \citep[e.g.][]{Sales2007,
  Cooper2010} according to the halo assebly history. We investigate in
Fig.~\ref{fig:satmass} the median satellite mass function of surviving
(solid) and merged (dashed) satellites for each of our
subsamples. Error bars correspond to dispersion obtained after 100
boostrap resampling of the data.

The surviving population in low and high $f_{\rm SH}$ is roughly
similar in shape, with a small excess of satellites for galaxies with
large stellar halo. {\it This weak but yet significant correlation
  between stellar halo content and observed number of satellites is an
  interesting prediction of our model that may become available to
  observations soon as a larger number of galaxies are observed to low
  surface brightness limits.}
  On the other hand, the spectrum of merged satellites shows instead a
  strong and significant difference in the shape of the mass
  distribution of merged satellites (see dashed curves in
  Fig.~\ref{fig:satmass}. {\it We find a large preference for centrals
    with low stellar halo fraction to have accreted an excess of low
    mass satellites and a deficit of high mass satellites compared to
    centrals with high stellar halo content}.

Counterintuitively perhaps, centrals with low stellar halo content
have had on average a {\it larger} number of mergers compared to the
high $f_{\rm SH}$ sample, albeit such mergers have been dominated by
satellites of smaller mass than in the high $f_{\rm SH}$
sample. Centrals with low stellar halo content have mostly
avoided mergers with satellites with stellar mass $M_{*,\rm sat} \sim
10^9 \; \rm M_\odot$ and above, whereas such events are not rare for
galaxies with similar virial mass but larger stellar halo
fractions. We conclude that the mass of the merged satellite and not
the number of mergers, is a dominant factor in determining the stellar
halo content of the central.

As the population of merged satellites largely determines the amount 
of stellar halo content, we investigate their
times of infall $t_{\rm inf}$, and find they differ
significantly. Infall time is defined as the time at which a satellite
reaches its peak total mass, a definition that roughly coincides with
the last time an object was considered a central before it accreted
onto a larger system.  The right panel of Fig.~\ref{fig:satmass} shows
an average weighted histogram with the infall times of merged
satellites, where each satellite is weighted by their stellar mass
(results do not change significantly if we plot the unweighted
histogram). As expected, low $f_{\rm SH}$ centrals have on average
less stellar mass brought in by satellites, but these satellites also
infall at earlier times than the high $f_{\rm SH}$ population. The
earlier infall times partially explains the excess of low mass
satellites that merged to centrals of low $f_{\rm SH}$, as galaxies that
infall at earlier times will have, at fixed total mass, a larger
fraction of their baryons in the form of gas compared to a satellite
that infalls later and have had the time to turn more of their gas into stars.

Moreover, the earlier infall times for merged satellites in centrals
with low stellar halo content are also consistent with the time of
rapid build up of the $M_{200}$ and stellar halos in these objects
according to Fig.~\ref{fig:mvst_all}. To guide the eye, the blue and
red vertical lines denote the $t_{50}$ of the low and high $f_{SH}$
populations, respectively.  This suggests that for the centrals with
low $f_{\rm SH}$ no significant satellite infall has occurred since $z
\sim 1$, consistent with the hypothesis of early assembled halos and
implying that not many phase-space features such as stellar streams or
substructures are expected in these hosts today. For high $f_{\rm SH}$
the incidence of young tidal features such as streams could be more
common.

\section{The in situ vs. accreted halos}
\label{sec:origin}

\begin{center}
\begin{figure}
\includegraphics[width=84mm]{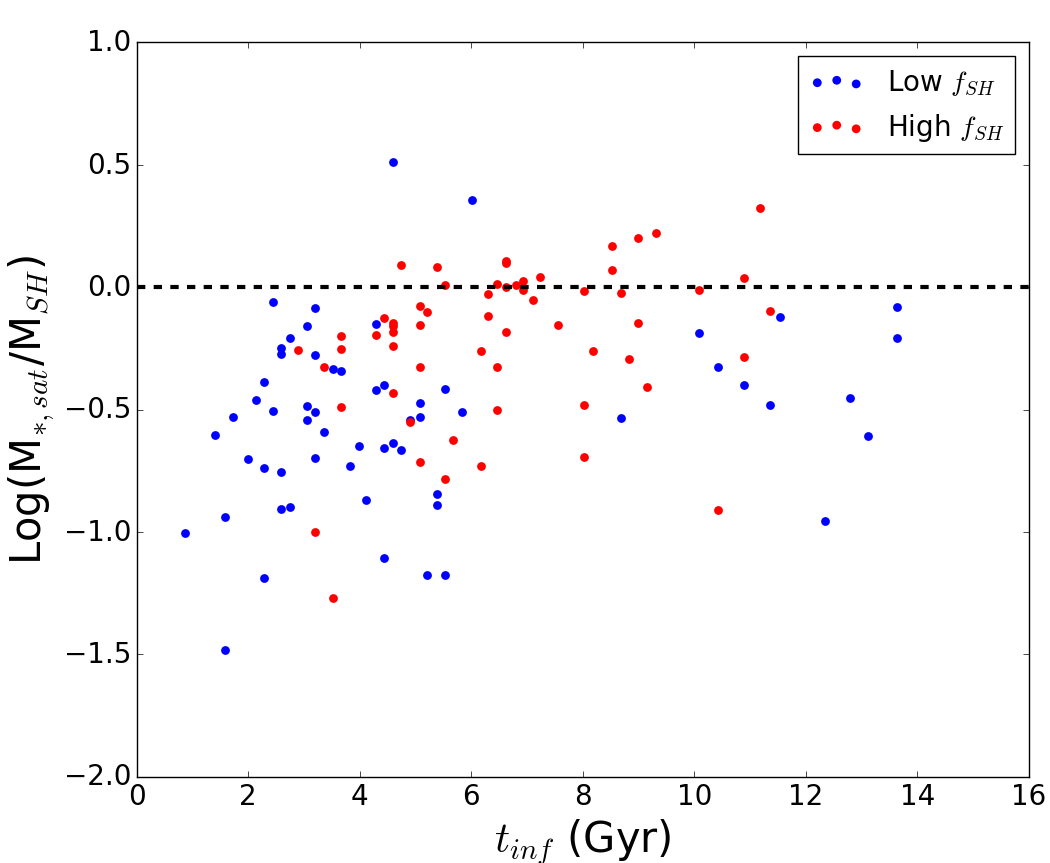}
\caption{Fraction of the total mass of the stellar halo that is
  contributed by the most massive accretion event as a function of the
  infall time of such an accreted satellite. In galaxies with high
  stellar halo fractions (red dots), the stellar halo mass is
  dominated by that single accretion event, as seen by the proximity
  to the horizontal line. For galaxies with low stellar halo fraction
  (blue dots) instead, the distribution is much wider, with cases
  where the most massive accretion event contributed less than $10\%$
  of the total halo mass, suggesting a fundamental difference in the
  build up of the stellar halos of these two samples.}
\label{fig:biggestsat}
\end{figure}
\end{center}

It is commonly assumed that the majority of the stars in stellar
halos were contributed by a single most massive event
\citep{DSouza2017, Deason2016}. We explore this for our sample in
Fig.~\ref{fig:biggestsat} where for each central in our low and high
$f_{\rm SH}$ subsample we select the most massive (in stars) merged
satellite over their entire history. For such an event, we record the
infall time and its maximum stellar mass, $M_{*,\rm sat}$. By
comparing $M_{*,\rm sat}$ to the mass of the stellar halo at $z=0$ we
compute the fraction of the stellar halo mass that could have been
brought in by the most massive merger event ($y$-axis). Note that this
is an upper limit since some of the stars could have been deposited
onto the central galaxy and not in the stellar halo itself (that is
why some of the points scatter above the 1 line). We find that most of
the objects sit near a ratio of 1, especially for the centrals with
high $f_{\rm SH}$. However, there are a large number of systems where
the fraction of stars in the most massive merger is only a minor
contributor, meaning that the stellar halo mass was either built by
more than one event or it has a completely different origin not
associated to accretion from satellites. 

Further exploration of the systems with $M_{*,\rm sat}/M_{SH}<0.1$
demonstrated that a large fraction of the stars were born in situ and
not accreted from satellites. This seems to occur more frequently in
our centrals with low fraction of stellar halos (blue dots in
Fig.~\ref{fig:biggestsat}) and earlier assembly (low $t_{\rm
  inf}$). In practice, we followed \citet{Rodriguez-Gomez2016} to
determine the origin of stellar particles, where a star is labeled
``in situ'' if it belonged to the main branch of the progenitor tree
at the time when it was born, or ``accreted'' if it was formed
in a different substructure.  The contribution of in situ stars to
stellar halos have been suggested before, both in simulations
\citep[e.g., ][]{Zolotov2010, Purcell2010,Font2011,Tissera2013} and
also in observations of the Milky Way, based for example on metallicity
and kinematics difference between the inner halo --presumed to be
formed in situ-- and the external stellar halo --mostly accreted--
\citep[e.g., ][]{Nissen2010,Carollo2007,Carollo2010}. In Illustris,
the in situ component of stellar halos seems significant for a large
number of objects, especially those with a relatively low fraction of
stellar halo. 

\begin{center}
\begin{figure*}
\begin{subfigure}{.475\textwidth}
\includegraphics[width=\textwidth]{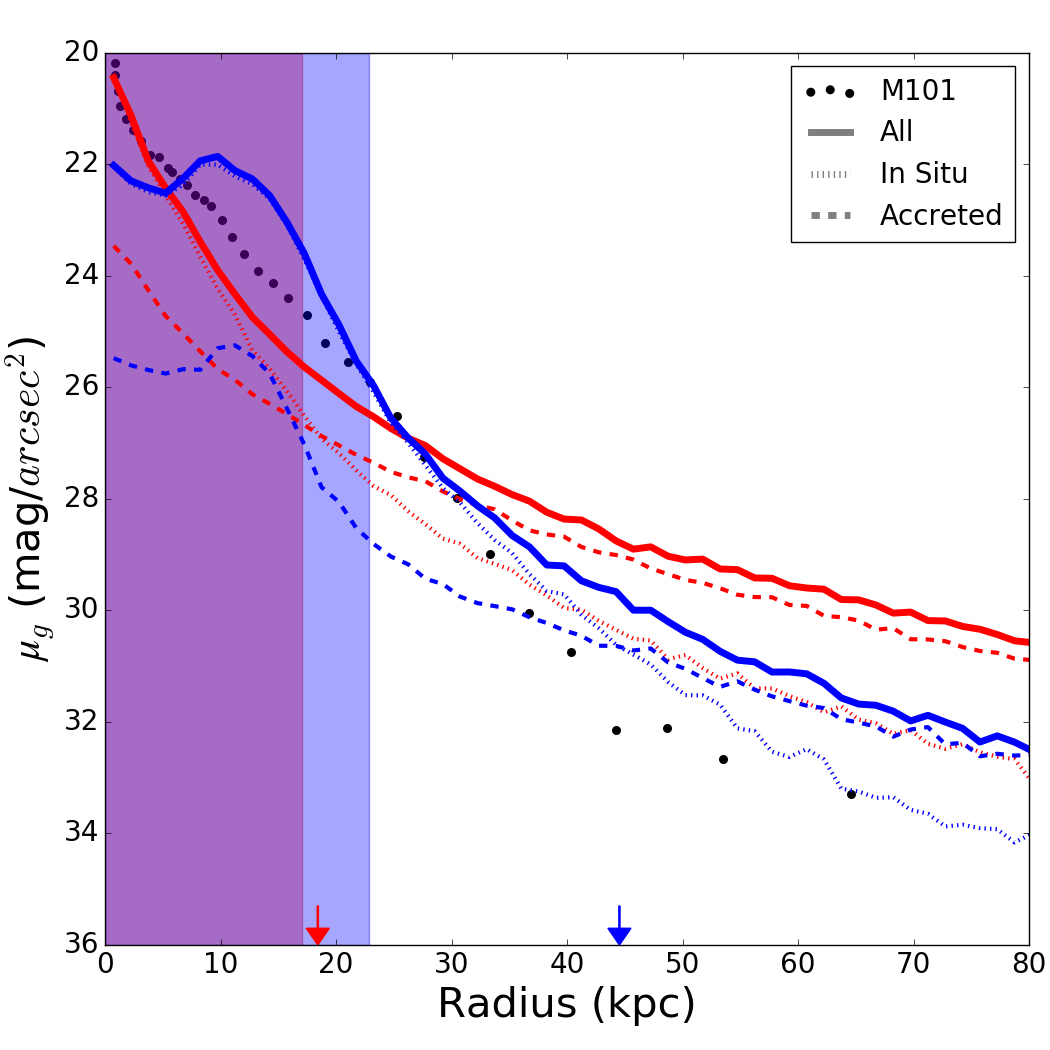}
\end{subfigure}
\begin{subfigure}{.475\textwidth}
\includegraphics[width=\textwidth]{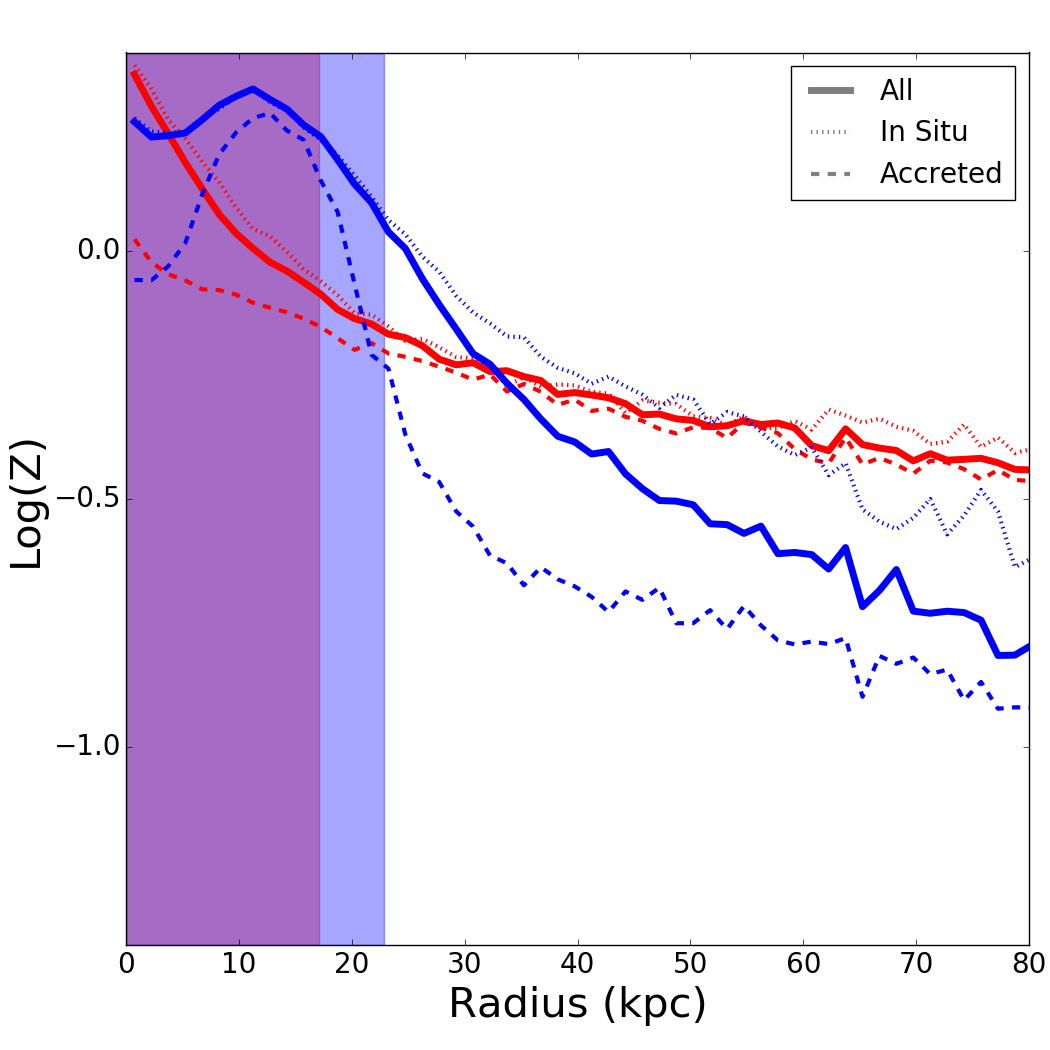}
\end{subfigure}
\caption{{\it Left}: Contributions from the in situ (dotted) and
  accreted (dashed) populations to the total surface brightness
  profile in the g-band (solid). The transition from in situ dominated
  to accreted dominated regime for galaxies with high $f_{\rm SH}$
  (red curves) 
  occurs roughly at the optical edge of the galaxy (green vertical
  region shows twice the average half-mass radius of the stars in the
  sample). In galaxies with low stellar halo fractions (blue lines)
  that transition takes place much further out, $r \sim 45$ kpc,
  indicating that in situ formed stellar halo strongly dominate the
  inner halos in these galaxies. {\it Right}: Median metallicity
  profiles expected for the stellar halos in our low and high $f_{\rm
    SH}$ sample. A much steeper radial gradient in the inner halo
  might be the smoking gun of a preferentially in situ-formed
  halo. }
\label{fig:allprofs}
\end{figure*}
\end{center}

This is shown more clearly in the left panel of
Fig.~\ref{fig:allprofs}, where we show the median surface brightness
profile for low and high $f_{\rm SH}$ subsamples (solid line)
separated into the in situ contribution (dotted line) and accreted
(dashed line). Here, we rotate each galaxy to be face-on (although our
results do not depend strongly on this) and add the $g$-band
luminosity of all stars contributing in each radial annulus. The
luminosity is then normalized by the area of each annulus and the
galaxies are assumed to be at a distance $7$ Mpc to convert to
apparent magnitude. Vertical shaded areas indicate the regions within
twice the average half mass radius of the stars in each subsample. By
definition, the stellar halo considered in this work extends outside
the indicated shaded rectangles.

In the case of centrals with low stellar halo content (blue curves),
the in situ population dominates the profile in the inner galaxy
region, as expected, but beyond the optical radius as well, closely following
 the solid (all stars) line out to very large distances. For
instance, one can consider the radius where the in situ and the
accreted population contribute equally to the total profile, indicated
by a blue vertical arrow, which reads $r\sim 45$ kpc in the case of
the low $f_{\rm SH}$ sample. Instead, for the galaxies with a massive
stellar halo (red curves), the transition radius occurs almost at the
edge of the galaxies, or $r \sim 18$ kpc, being accretion-dominated
throughout most of the stellar halo component.

For comparison, we include in the left panel of
Fig.~\ref{fig:allprofs} the observed density profile of M101 from
\citet{Merritt2016}, which seems to agree well with the average in
situ component in the low $f_{\rm SH}$ in Illustris. Although our
simulations do not have an analog with such a low stellar halo
component as measured for M101 (see Fig.~\ref{fig:obs}), it is
interesting to see that the in situ distribution of stars predicts
such a similar overall shape for this diffuse component. If Illustris
indeed overestimates the accreted contribution due to the larger
number of dwarf galaxies predicted with respect to observations, {\it
  there is some room to suggest that the halo of M101 and galaxies
  alike might be the ideal place to study this elusive in situ
  component in stellar halos}.

The different origins predicted for the stellar halos of galaxies with
low and high fraction of stars on their halos are also evident in the
metallicity profile shapes. Right panel of Fig.~\ref{fig:allprofs}
shows the metallicity profile predicted for all stars (solid) and the
in situ (dotted) and accreted (dashed) components. As expected, the in
situ stars have a larger metallicity compared to the accreted
component since they were born in a more massive progenitor. However,
for galaxies in the low $f_{\rm SH}$ sample there is an intrinsic
steep negative slope in the change of metallicity with radius and
since it dominates out to large radii, the overall metallicity profile
changes rapidly as we move outwards. By contrast, in the high $f_{\rm
  SH}$ sample, the accreted component dominates and has a more shallow
slope as a product of the mixing of material of stripped satellites at
different radii. This agrees with and extends previous results on
stellar halos of early types galaxies presented in \citet{Cook2016}.
{\it Our results indicate that a steep metallicity gradient may be the
  smoking gun of a mostly in situ formed stellar halo}. Such profiles
are expected to be more common around galaxies with low fraction of
mass on their stellar halos, but the signature is expected to be
independent of mass content.

\section{Summary and Discussion}
\label{sec:conc}

We use the Illustris cosmological hydrodynamical simulations to study
the assembly of stellar halos in galaxies of mass comparable to the MW
within the $\Lambda$CDM scenario. In particular, for a narrow range of
virial halo mass $M_{200}=8 \times 10^{11}$-$2\times10^{12} \; \rm M_\odot$ we focus on the relative
amount of stellar mass in the extended halos normalized to the stars
in the central galaxy, $f_{\rm SH} = M_{*,\rm SH}/M_{*,\rm
  gal}$. Stellar halo mass is calculated as all stellar mass within
the virial radius that is beyond twice the half mass radius of the
stars (considered the galaxy) and is not associated to satellites. Our
results can be summarized as follows:

\begin{itemize}
\item Simulated stellar halos span a large range in $f_{\rm SH} \sim
  5$-$50\%$ with a median value $ \langle f_{\rm SH} \rangle \sim 21\%$. Stellar
  halo masses for this virial mass range can therefore be as low as
  $M_{*,\rm SH}=2 \times 10^{9} \; \rm M_\odot$ or as massive as $6 \times 10^{10} \; \rm
  M_\odot$. Simulated halos can be fairly oblate, with median
  $\langle b/a \rangle \sim 0.87$ and $\langle c/a \rangle \sim 0.49$, where $a,b,c$ are the major to
  minor principal axes of the inertia tensor for the stars. This
  departure from sphericity can generate significant differences in
  the derived surface brightness profiles of the halos, with face-on
  objects being up to two magnitudes fainter than the same objects
  when viewed edge-on. Caution should therefore be exercised when
  analyzing observed profiles of galaxies at different inclination
  angles.

\item Simulations present a large variety of stellar halo profiles and
  masses which can accommodate the diversity of observed stellar halos 
  to a certain degree. However, when considering the specific
  shape of the profiles, simulated galaxies in Illustris are in tension with
   the most steeply declining surface
  brightness profiles such as those observed in NGC1042 and NGC3351.
  The excess of stars at large distances in the
  simulations can partially be attributed to the overly massive low mass end of
  the mass function, meaning that merging satellites contribute more
  stars than they should to the stellar halos. Yet it is 
  encouraging that galaxy formation models are able to predict a set of 
  quite distinct behaviours for the surface brightness profiles of extended
  halos, some with relatively shallow profiles whereas other galaxies
  in the same mass range might show a rather rapidly declining stellar
  light in their outskirts \citep[see also ][]{Pillepich2014}.

\item We study the origin of this diversity of stellar halos by
  selecting the $5\%$ tails of low and high stellar halo fractions,
  $f_{\rm SH}$. We find fundamental differences in the properties of
  the central galaxies in each subsample, with objects with low
  stellar halo content (low $f_{\rm SH}$) being in general more
  disk-dominated and star-forming whereas high $f_{\rm SH}$ centrals
  are spheroidals and not forming stars. On the other hand,
  environment seems not to play a major role in the assembly of the
  stellar halo, as the present-day density of galaxies around
  centrals, as characterized by $\Sigma_5$, of low and high $f_{\rm
    SH}$ galaxies are remarkably similar. These findings are in good
  agreement with observations.

\item Galaxies with low stellar halo content form in halos that
  assemble more rapidly, reaching half their present day mass on
  average by $z_{50}=1.5$. In contrast, galaxies with a massive stellar
  halo are more typical in late-forming halos, with a median redshift
  for the assembly of half their mass $z_{50}=0.73$. The scatter from
  object to object is, however, large. Similarly, satellites that get
  tidally disrupted building the stellar halos infall earlier and are
  less massive in galaxies with low mass stellar halos. Average infall
  times for low and high stellar halo centrals are $<t_{\inf}>=2.6$ and
  $5.2$ Gyr, respectively.

\item In Illustris, stellar halos can have a significant fraction of in
  situ formed stars. Those stars form over time within $\sim 30$
  kpc of the center, but are located today out to $\sim 100$ kpc,
  favoring a dynamical scenario for their launch into their more
  extended orbits. In situ formed halos decline steeply with radius
  and can dominate the contribution in low $f_{\rm SH}$ objects. We
  argue therefore that extreme objects such as M101 and the lowest stellar halo
  galaxies are the prime candidates to observationally study the in
  situ formed stellar halos. Furthermore, we predict that stellar
  halos formed mainly by contribution of in situ stars should have a
  markedly steep metallicity profile that can be used as a 'smoking gun'
  for identification in observations.
\end{itemize}

The diversity of stellar halo shapes is an interesting new avenue to
recover information on the past merger history of galaxies through
their present day properties. Numerical simulations seem to naturally
reproduce a large variety of shapes and mass for stellar halos in the
Milky Way mass regime. Whether the simulated diversity matches 
observations or not is not yet settled. To first order, galaxies with little
or no stellar halo are difficult to find in cosmological simulations
within $\Lambda$CDM where mergers are a prevalent feature. Biases 
and sky subtractions in observations that push the boundary of
observability are challenging and difficult to quantify, which, in addition to the
scarcity of objects observed to very faint surface brightness, hinders a
correct interpretation of the comparison between observations and
simulations. Current efforts such as {\sc GHOSTS}
\citep{Radburn-Smith2011} and Dragonfly observations
\citep{Merritt2016} have been transformative. But more complete
samples of galaxies at a fixed mass and varied morphology will lay the
ground for a more fair comparison with models.  From this perspective,
projects targeting the faint outskirts of galaxies in large numbers
are a necessity. The HERON Survey \citep[Halos and Environments of
Nearby Galaxies, ][]{Rich2017} and future missions such as WFIRST might
help define the expected global properties of stellar halos and
whether theoretical predictions are (or not) a good match to the
observed Universe.

\section{Acknowledgments}

We would like to thank Mario Abadi, Richard D'Souza, Benedikt Diemer and Annalisa
Pillepich for their valuable input and feedback on the results. LVS
acknowledges financial support from the Hellman Foundation.

\bibliography{master}

\end{document}